\documentclass[namedreferences]{solarphysics}

\usepackage[optionalrh]{spr-sola-addons} 
\usepackage[]{graphicx}        
\usepackage{rotating}
\usepackage[labelsep=space]{caption}
\usepackage{color}           
\usepackage{url}             
\usepackage[breaklinks=true]{hyperref}
\usepackage{mynotes}

\newcommand{{\kms}}{    {$\mathrm {km\,s^{-1}}$}}
\newcommand{{\ms}}{    {$\mathrm {m\,s^{-1}}$}}

\begin{document}
\def\mntnoteheading{}
\def\mntnoteformat{\parindent=0em \leavevmode \llap{\makemntmark}}

\begin{article}

\begin{opening}

\title{The Horizontal Component of Photospheric Plasma Flows During the
Emergence of Active Regions on the Sun}

\author{A.~\surname{Khlystova}$^{1}$
       }
\runningauthor{A. Khlystova}
\runningtitle{The Horizontal Photospheric Flows During the Emergence of Active
Regions}

   \institute{$^{1}$ The Institute of Solar-Terrestrial Physics, Siberian
Branch, Russian Academy of Sciencess, Irkutsk, Russia\\
                     email: \url{hlystova@iszf.irk.ru}
             }

\begin{abstract}
The dynamics of horizontal plasma flows during the first hours of the emergence
of active region magnetic flux in the solar photosphere have been analyzed using
SOHO/MDI data. Four active regions emerging near the solar limb have been
considered. It has been found that extended regions of Doppler velocities with
different signs are formed in the first hours of the magnetic flux emergence in
the horizontal velocity field. The flows observed are directly connected with
the emerging magnetic flux; they form at the beginning of the emergence of
active regions and are present for a few hours. The Doppler velocities of flows
observed increase gradually and reach their peak values 4\,--\,12 hours after
the start of the magnetic flux emergence. The peak values of the mean (inside
the $\pm$500\,{\ms} isolines) and maximum Doppler velocities are
800\,--\,970\,{\ms} and 1410\,--\,1700\,{\ms}, respectively. The Doppler
velocities observed substantially exceed the separation velocities of the
photospheric magnetic flux outer boundaries. The asymmetry was detected between
velocity structures of leading and following polarities. Doppler velocity
structures located in a region of leading magnetic polarity are more powerful
and exist longer than those in regions of following polarity. The Doppler
velocity asymmetry between the velocity structures of opposite sign reaches its
peak values soon after the emergence begins and then gradually drops within
7\,--\,12 hours. The peak values of asymmetry for the mean and maximal Doppler
velocities reach 240\,--\,460\,{\ms} and 710\,--\,940\,{\ms}, respectively. An
interpretation of the observable flow of photospheric plasma is given.
\end{abstract}

\end{opening}

\section{Introduction}
     \label{sec:intro}
\par According to measurements of Doppler velocities during the emergence of
active regions in the solar photosphere, the presence of upflows at the tops
\cite{bra85a,bra85b,tar89,lit98,str99,kub03,gug06,gri07,gri09} and downflows of
plasma at the footpoints
\cite{gop67,gop69,kaw76,bac78,zwa85,bra85a,bra85b,bra85c,lit98,sol03,lag07,xu10}
of the emerging magnetic loops is well-established.

\par Uptil now horizontal photospheric velocities accompanying the emergence of
active regions have only been measured indirectly. \inlinecite{fra72} studied a
young active region and found that the photospheric magnetic field knots (21
events) moved along the arch filament system (AFS) at velocities of
0.1\,--\,0.4\,{\kms}. Using observations of the young active region for 6.5
hours, \inlinecite{sch73} found that magnetic elements moved in random
directions with velocities of 0.4\,--\,1.0\,{\kms}. \inlinecite{bar90} measured
the separation velocities of the opposite polarities for 45 bipolar pairs in the
young active region and obtained velocity values up to 0.5\,--\,3.5\,{\kms},
decreasing with time. \inlinecite{str99} considered the emerging magnetic flux
in a growing active region and found that the footpoints of the magnetic loops
separated at an average velocity of 1.4\,{\kms}. \inlinecite{gri09} estimated
the separation velocities of the photospheric magnetic flux outer boundaries in
the NOAA 10488 active region. The velocities decreased as the magnetic flux
emerged: they were 2\,--\,2.5\,{\kms} at the end of the first hour and
0.3\,{\kms} in two and a half hours.

\par High values of horizontal photospheric velocities were obtained during the
emergence of ephemeral active regions, except for the work of \inlinecite{cho87}.
\inlinecite{har73} found that during the first 30 minutes of the emergence of
ephemeral active regions the footpoints of magnetic loops separated at
5\,{\kms}; the divergence velocity decreased to 0.7\,--\,1.3\,{\kms} over the
next six hours and continued to decrease later on. \inlinecite{cho87} obtained
low separation velocities of opposite polarities of 0.2\,--\,1\,{\kms} for 24
emerging ephemeral active regions. \inlinecite{hag01} found that the outer
boundaries of the ephemeral regions expand with velocities up to 5.5\,{\kms};
there is a tendency for the velocities to decrease with time.

\par \inlinecite{ots11} studied 101 emerging flux regions of different spatial
scales. They found that the separation velocities of opposite polarities are
lower than 1\,{\kms} for large emerging magnetic fluxes, whereas they reach
4\,{\kms} in small-scale ones.

\par Interesting results were obtained from the analysis of horizontal flows in
the emerging active regions by granular motion. \inlinecite{str96} found
large-scale horizontal divergent granular flows in a growing active region which
were comparable with the common drift of magnetic polarities. The authors
interpreted this fact as a close interaction between granulation and magnetic
fields. \inlinecite{koz05} and \inlinecite{koz06} also found divergent flows
located between the footpoints of emerging flux loops. These authors consider
these flows to be either convective flows which may trigger magnetic flux
emergence from deep layers. They also suggest that the observed flows are formed
by the emergence of magnetic flux. Note that the articles listed above concerned
developing active regions already containing pores.

\par The present investigation involves an analysis of photospheric
Doppler velocities for active regions emerging near the limb. This subject was
considered earlier by \inlinecite{khl11}.

\section{Data Analysis}
     \label{sec:data}

\par We used full-disk solar magnetograms and Dopplergrams in the photospheric
line Ni\,{\sc i} 6768\,\AA\ and continuum images obtained on board the
\textit{Solar and Heliospheric Observatory} (SOHO) using the \textit{Michelson
Doppler Imager} (MDI) \cite{sch95}. The temporal resolution of the magnetograms
and Dopplergrams is 1 minute, while that of the continuum is 96 minutes. The
spatial resolution of the data is 4$''$, and the pixel size is approximately
2$''$.

\par We have cropped a region of emerging magnetic flux from a time sequence of
data, taking into account its displacement caused by solar rotation. The
approximate displacement of the region was calculated by the differential
rotation law for photospheric magnetic fields \cite{sno83}. The exact tracking
was performed by applying a cross-correlation analysis to two magnetograms
adjacent in time. A precise spatial superposition of the data was achieved by
cropping fragments with identical coordinates from simultaneously acquired
magnetograms, Dopplergrams, and continuum images. For the Dopplergrams we
applied a moving average over five images to reduce a contribution of
five-minute oscillations to the velocity signal. The solar differential rotation
and other factors distorting the Doppler velocity signal were removed by using
the following technique. We averaged three upper and three lower rows of the
cropped region and linearly smoothed the averaged rows, thus obtaining upper and
lower rows of the array of solar rotation velocities. To obtain the internal
field of the array, we performed a linear interpolation between the upper and
lower pixels of the columns belonging to the smoothed rows. The obtained array
of solar rotation velocities was subtracted from the original one. 

\par The parameters under study were calculated in the region of the emerging
magnetic flux. The boundary of the emerging flux region was visually
inspected. For active regions we determined the following.
\begin{itemize}
 \item $\theta$ is the heliocentric angle characterizing the distance from the
solar disk center to the location of active region emergence; it is also
approximately the angle between the normal to the surface and the line of sight
to the emerging magnetic flux:
\begin{eqnarray}
\theta = \arcsin(r/R),
\end{eqnarray}
where $r$ is the distance from the solar disk center to the location of active
region emergence and $R$ is the solar radius.
 \item $\Phi_{max}$ is the total unsigned magnetic flux at the maximum
development of the active region which was measured inside isolines $\pm$60\,G,
taking into account the projection effect and supposing that the magnetic
field vector is perpendicular to the solar surface:
\begin{eqnarray}
\Phi_{max} = S_{0} \sum_{i=1}^N \frac{|B_{i}|}{cos \theta_{i}},\
\end{eqnarray}
where $N$ is the number of pixels with $|B_{i}|$ $>$ 60\,G, $S_{0}$ is the area
of the solar surface of the pixel in the center of the solar disk in $cm^{2}$, $B_{i}$
is the line of sight magnetic field strength of the $i-th$ pixel in $G$, and
$\theta_{i}$ is the heliocentric angle of the $i-th$ pixel. 
 \item {\it d$\Phi$/dt} is the total unsigned magnetic flux growth rate in the
first 12 hours of the active region emergence. 
 \item $V_{mean-}$ and $V_{mean+}$ are the peak values of the mean negative and
positive Doppler velocities inside the isoline, $-$500 and $+$500\,{\ms}, during
the period considered. The isoline of 500\,{\ms} was selected because it
outlines the observable Doppler velocity structures well and the main
contribution of convection flow is below this level. The Doppler velocity values
did not always exceed 500\,{\ms} in the calculation region, so the mean
velocities were not determined for these instants. 
 \item $V_{max-}$ and $V_{max+}$ are the peak values of absolute maximum
negative and positive Doppler velocities during the period considered. 
 \item $V_{sep}$ is a mean relative separation velocity between the two outer
boundaries of the photospheric magnetic flux of opposite polarities. It is
calculated over a two-hour period at the peak values of the Doppler velocities,
taking into account the projection effect:
\begin{eqnarray}
\overrightarrow{V}_{sep} = \frac{1}{2 (T_{2}-T_{1})} \left( \frac{L_{2}}{cos \theta_{2}} - \frac{L_{1}}{cos \theta_{1}} \right),
\end{eqnarray}
where $T_{2}-T_{1}$ is period of time under consideration in $s$, $L_{1}$ and
$L_{2}$ are the distances between the two outer boundaries of the photospheric
magnetic flux in the image plane for points of time $T_{1}$ and $T_{2}$ in $m$,
and $\theta_{1}$ and $\theta_{2}$ are the heliocentric angles corresponding to
the active region position in points of time $T_{1}$ and $T_{2}$. $V_{sep}$
contains the contribution of the horizontal expansion velocities of the emerging
magnetic flux $V_{exp}$ and the magnetic polarity displacement due to the
geometry of the rising magnetic loop $V_{foot}$. The contribution of the
magnetic flux expansion $V_{exp}$ will essentially exceed the displacement of
the magnetic polarities $V_{foot}$ at the very beginning of the active region
appearance. The displacement of the magnetic polarities resulting from vertical
emergence of the magnetic loop $V_{foot}$ will not give a Doppler shift. Thus,
the contribution of the magnetic flux expansion $V_{exp}$ to the Doppler
velocity signal will be calculated as
\begin{eqnarray}
V_{exp} \lesssim \overrightarrow{V}_{sep} sin \theta,
\end{eqnarray}
\end{itemize}

\begin{table}
\begin{tabular}{ccccrc}
\hline
{\bf Active}&								
{\bf Date\tabnote{Time at the start of active region emergence}}&  	
{\bf Coordinates\tabnote{Coordinates at the start of active region emergence}}& 
{\bf $\bf \theta$}&							
{\bf $\bf \Phi_{max}$,}&						
{\bf {\it \bf d$\Phi$/dt},}						
\\
{\bf regions}&								
{\bf }&								
{\bf }&									
{\bf }&									
{\bf Mx}&								
{\bf Mx\,h$^{-1}$}							
\\ \hline
9037 & 10 Jun 2000 -- 06:08 UT & N21E59 B$_{0}+$0.4 & 61$^\circ$ & $ 1.39\times10^{22}$ & $ 1.17\times 10^{20}$ \\
8536 & 06 May 1999 -- 00:51 UT & S24E65 B$_{0}-$3.7 & 66$^\circ$ & $ 1.88\times10^{22}$ & $ 1.13\times 10^{20}$ \\
8635 & 14 Jul 1999 -- 12:13 UT & N42W47 B$_{0}+$4.2 & 57$^\circ$ & $>2.97\times10^{21}$ & $ 9.50\times 10^{19}$ \\
9064 & 26 Jun 2000 -- 11:16 UT & S21W46 B$_{0}+$2.4 & 51$^\circ$ & $>3.61\times10^{21}$ & $ 1.63\times 10^{20}$
\\ \hline
\end{tabular}
\caption[]{Active regions studied}
\label{tbl:ar}
\end{table}

\section{The Investigated Active Regions}
 
\par
\begin{sidewaysfigure}
\centerline{
\includegraphics[width=\textheight]{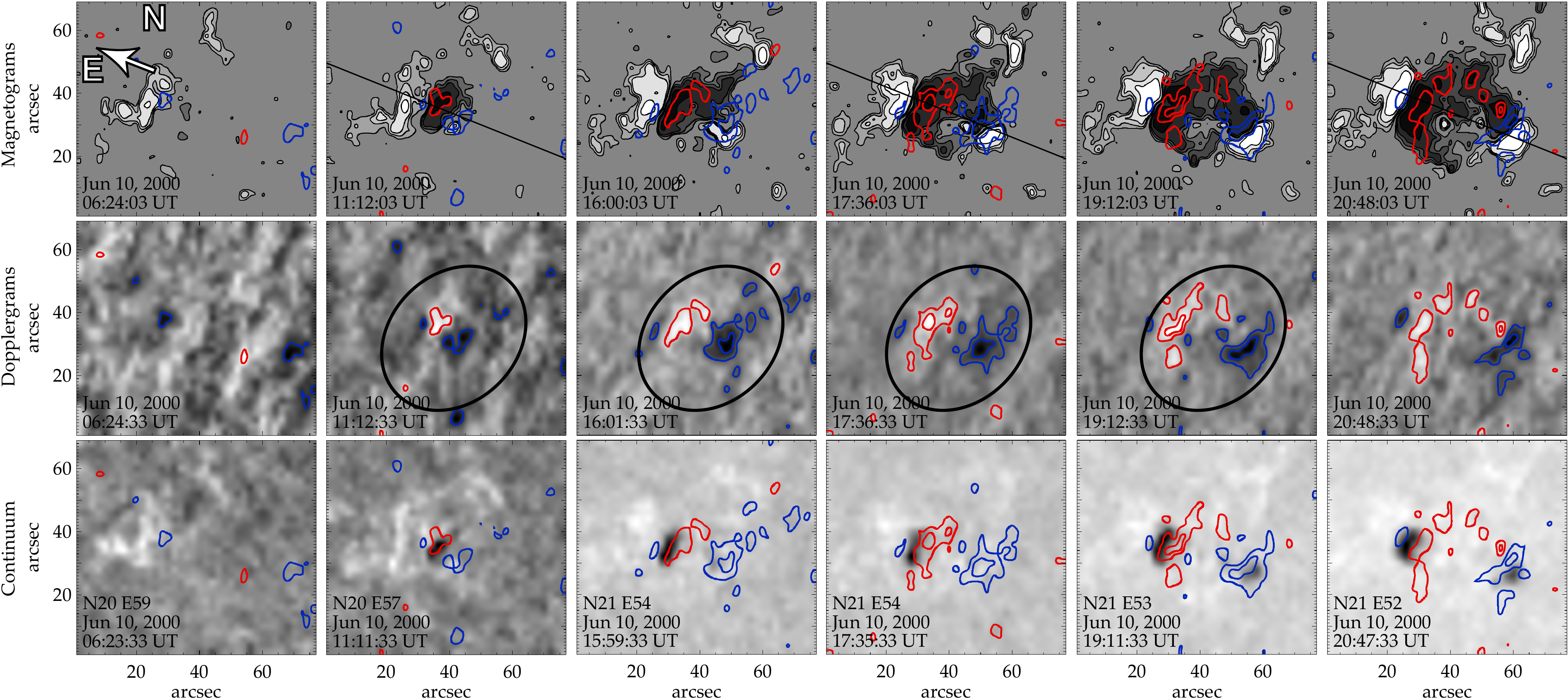}
}
\caption{
The active region NOAA 9037 emerges on 10 Jun 2000 at N21 E59. On the
magnetograms (isolines $\pm$60, 100, 150, 300\,G), Dopplergrams, and continuum
the isolines of Doppler velocities are superimposed. The blue isoline
corresponds to $-$500,$-$1000\,{\ms} -- plasma motion towards the observer; the
red isoline corresponds to $+$500,$+$1000\,{\ms} -- plasma motion away from the
observer. Forming Doppler velocity structures are marked by an ellipse. The
orientation of the images is shown in the upper left corner; the direction from
the solar disk center to the emerging magnetic flux is marked by a white arrow.
The black straight line on the upper-row panels marks the location of the slice
of the time slice diagrams.
	}
\label{fig:01}
\end{sidewaysfigure}

\begin{sidewaysfigure}
\centerline{
\includegraphics[width=\textheight]{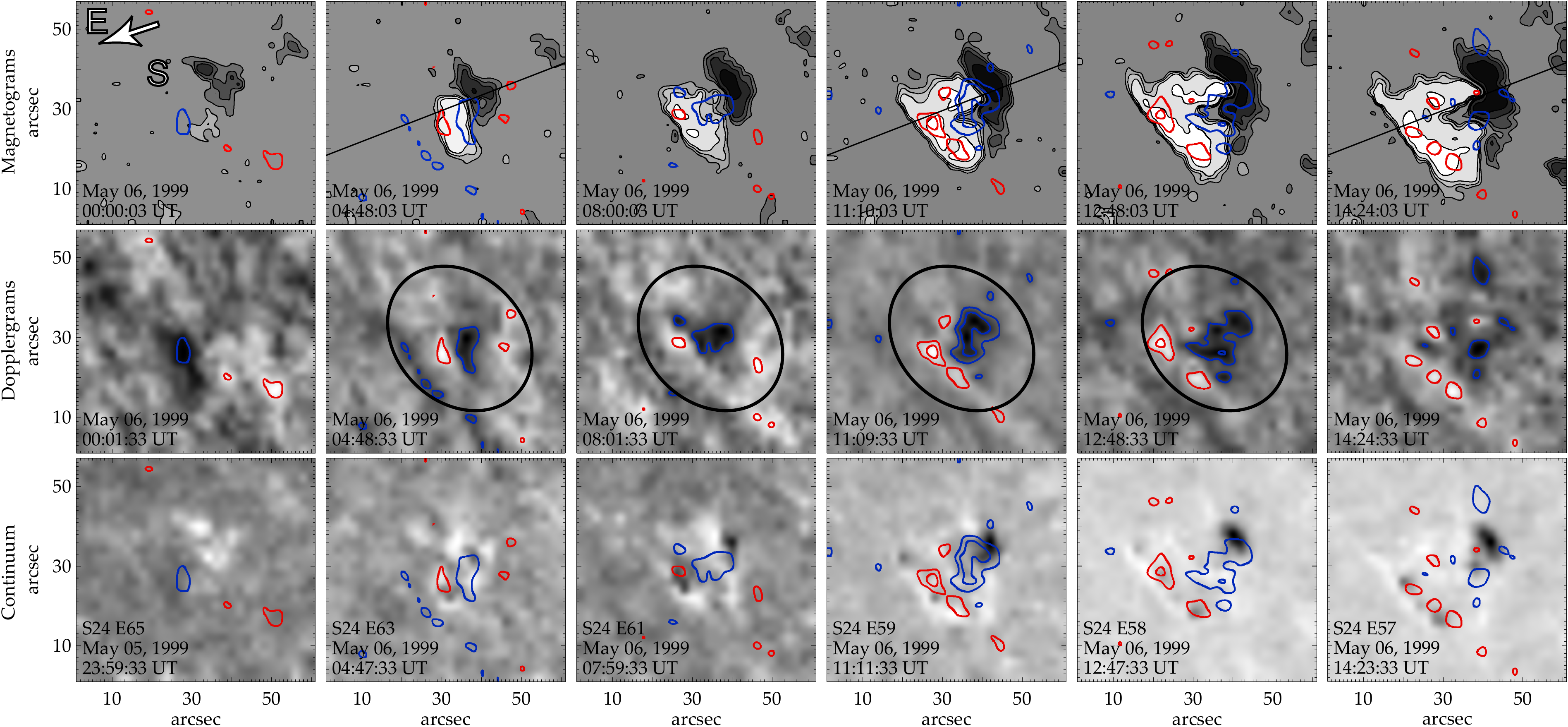}
}
\caption{
Active region NOAA 8536 emerges on 6 May 1999 at S24 E65. The marking convention
is the same as in Figure~\ref{fig:01}.
	}
\label{fig:02}
\end{sidewaysfigure}

\begin{sidewaysfigure}
\centerline{
\includegraphics[width=\textheight]{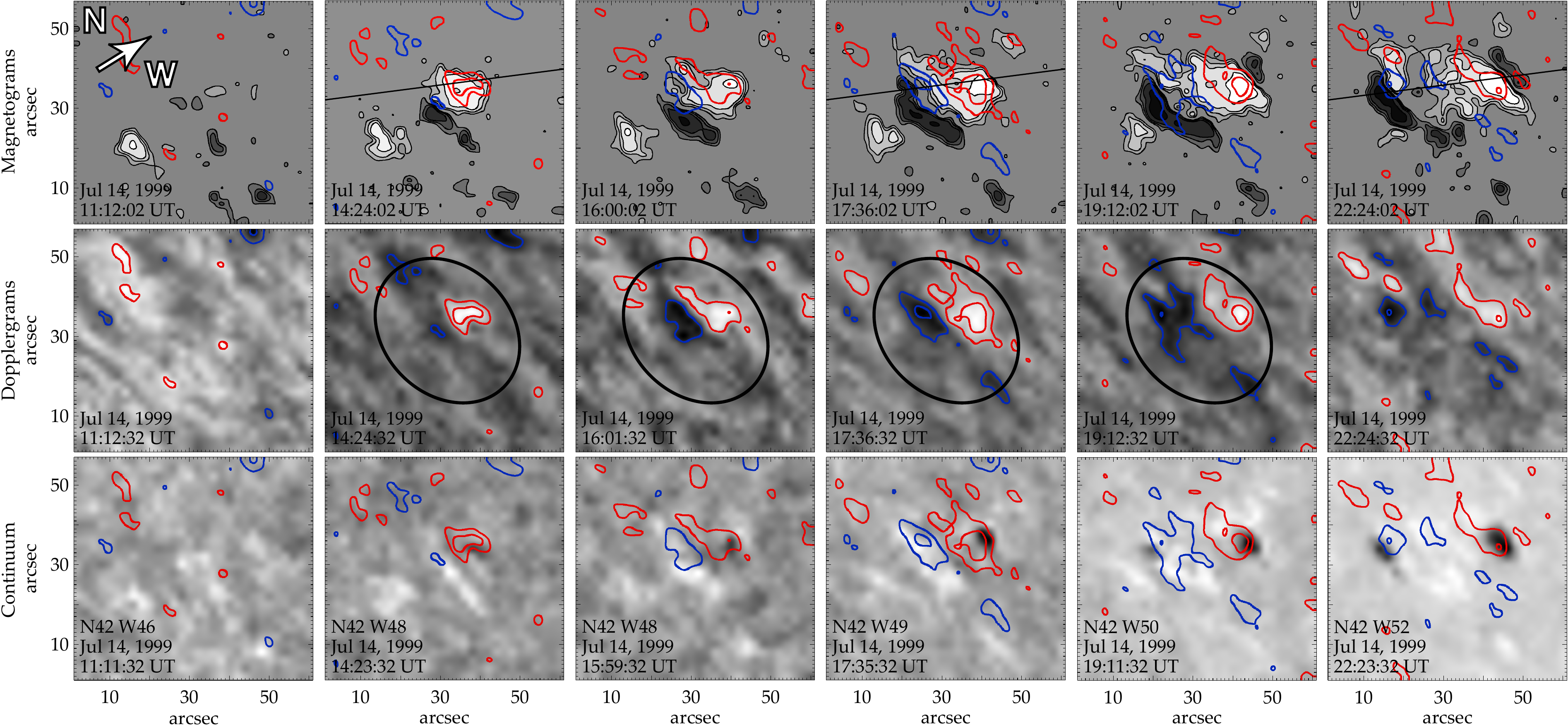}
}
\caption{
Active region NOAA 8635 emerges on 14 Jul 1999 at N42 W47. The marking
convention is the same as in Figure~\ref{fig:01}.
	}
\label{fig:03}
\end{sidewaysfigure}

\begin{sidewaysfigure}
\centerline{
\includegraphics[width=\textheight]{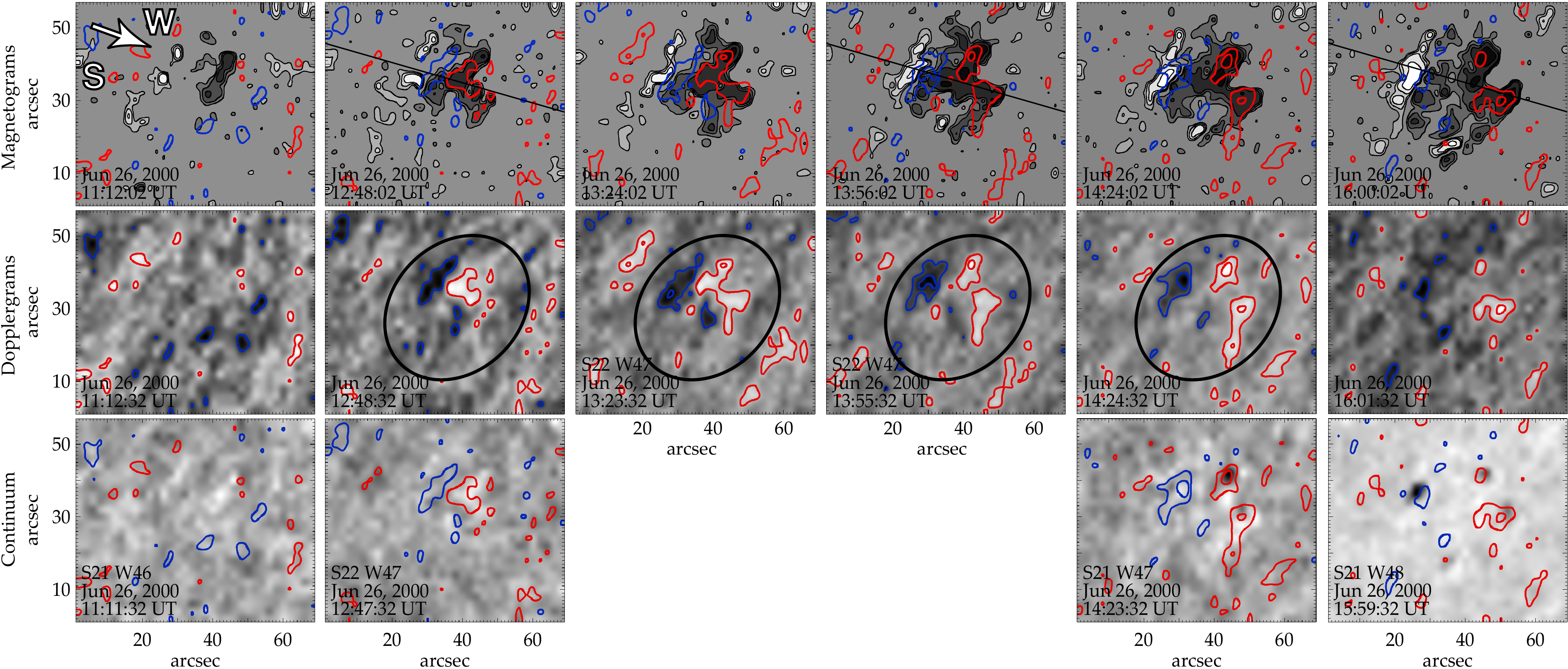}
}
\caption{
Active region NOAA 9064 emerges on 26 Jun 2000 at S21 W46. The marking
convention is the same as in Figure~\ref{fig:01}.
	}
\label{fig:04}
\end{sidewaysfigure}

 Four active regions emerging near the solar limb have been considered (see
Figures~\ref{fig:01}, \ref{fig:02}, \ref{fig:03}, and \ref{fig:04}). The continuum
images show that in the first hours of the emergence of active regions there are
only pores. In the first hours of appearance, active regions have a high
magnetic flux growth rate (Table~\ref{tbl:ar}).

\par Because of the projection effect of the magnetic field vector to the line
of sight, the emergence of active regions begins with the appearance of one,
then another magnetic polarity. The boundary where the magnetic field changes
sign is not the polarity inversion line (the upper-row panel in
Figures~\ref{fig:01}, \ref{fig:02}, \ref{fig:03}, and \ref{fig:04}). The
polarity inversion line location can be indirectly estimated by the pore
positions. They arise in both polarities of the studied active regions (the
bottom-row panel in Figures~\ref{fig:01}, \ref{fig:02}, \ref{fig:03}, and
\ref{fig:04}). The polarity inversion line will pass in the middle between the
leading and following pores with a small shift towards the following pore at the
beginning of the magnetic flux emergence due to the geometrical asymmetry of the
emerging magnetic flux (Figure 4a of \opencite{dri90}). Thus, the observable
boundary where the magnetic field changes sign lies in the polarity located
closer to the solar disk center. In the active regions, the axis connecting
opposite magnetic polarities rotates as the magnetic flux emerges (for an
in-depth analysis see \opencite{luo11}).

\section{Doppler Velocity Structures}

\par The SOHO/MDI data have a low spatial resolution which enables one to
observe large-scale flows of plasma. The active regions under consideration
emerge in different sectors of the solar disk, but the morphology of Doppler
velocity structures during the first hours of magnetic flux emergence is
similar. An amplification of negative Doppler velocities (plasma motion towards
the observer) is observed on the boundary where the magnetic field changes sign
located at the disk-side polarity (closer to the solar disk center) because of
the projection effect of the magnetic field vector to the line of sight and of
positive Doppler velocities (plasma motion away from the observer) in the
limb-side polarity (closer to the solar limb) (Figures~\ref{fig:01},
\ref{fig:02}, \ref{fig:03}, and \ref{fig:04}). High Doppler velocities inside
the $\pm$1000\,{\ms} isolines occupy significant areas (up to 50\% of the
Doppler velocity structure area inside the 500\,{\ms} isoline within individual
time intervals) that are localized in the central part of Doppler velocity
structure and exist for a long time: for NOAA 9037 on June 10 at 17:36 and 19:12
UT (Figure~\ref{fig:01}); for NOAA 8536 on May 6 at 11:09 UT
(Figure~\ref{fig:02}); for NOAA 8635 on July 14 at 14:24 and 17:36 UT
(Figure~\ref{fig:03}); for NOAA 9064 on June 26 at 13:55 and 14:24 UT
(Figure~\ref{fig:04}).

\par
\begin{figure}
\centerline{
\includegraphics[width=11.53cm]{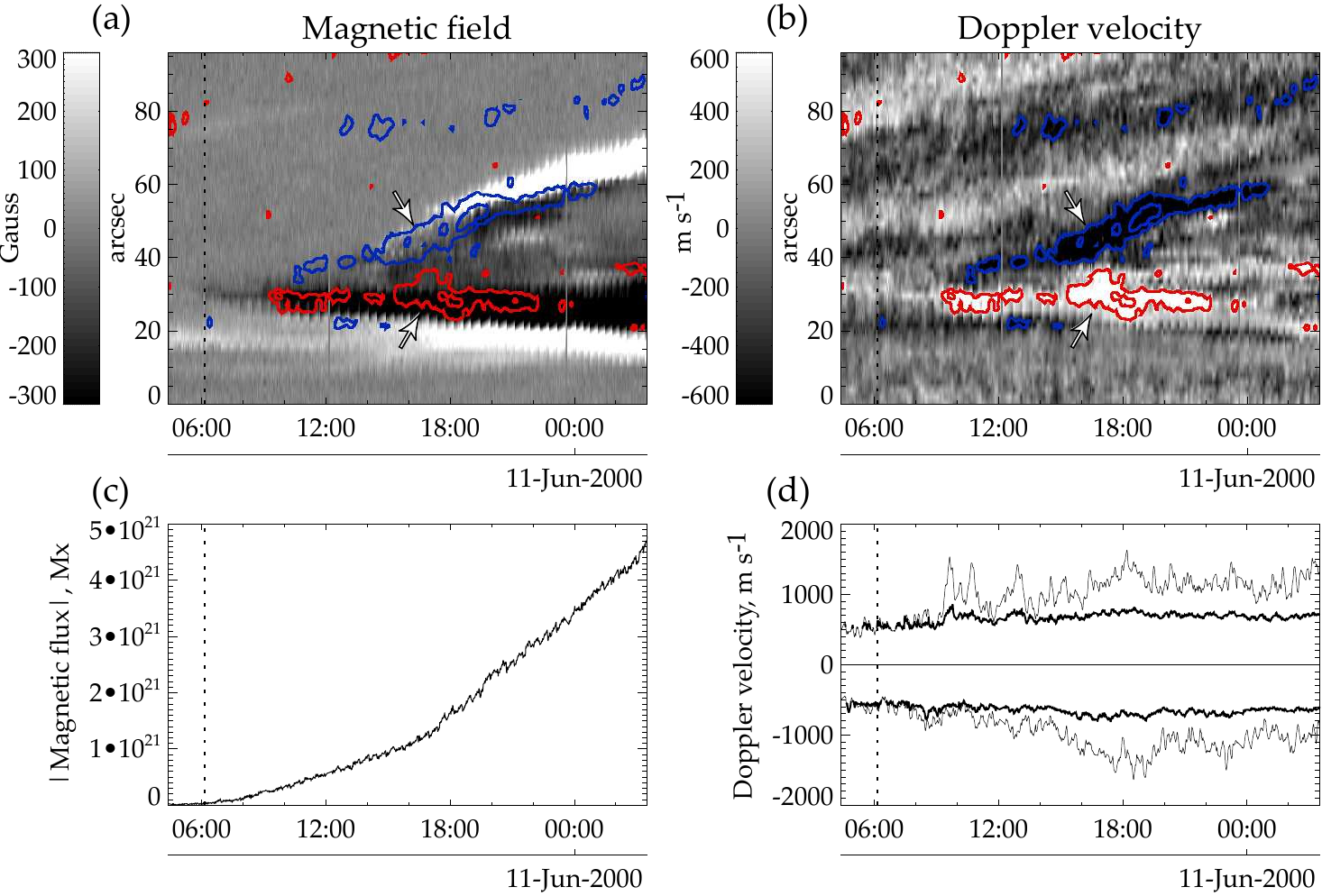}
}
\caption{
Active region NOAA 9037: time slice diagrams of (a) magnetic field strength and
(b) Doppler velocities. The location of the slice is marked on the magnetograms
in Figure~\ref{fig:01} by a black line. The blue and red isolines correspond to
$-$500, $-$1000 and $+$500, $+$1000\,{\ms}, and the Doppler velocity structures
analyzed are marked by arrows; (c) time variation of the total unsigned magnetic
flux; (d) time variation of mean (thick line) and absolut maximum (thin line)
values of negative and positive Doppler velocities in the region of emerging
magnetic flux. The vertical dotted line marks the time of the beginning of
magnetic flux emergence.
	}
\label{fig:05}
\end{figure}

\begin{figure}
\centerline{
\includegraphics[width=11.53cm]{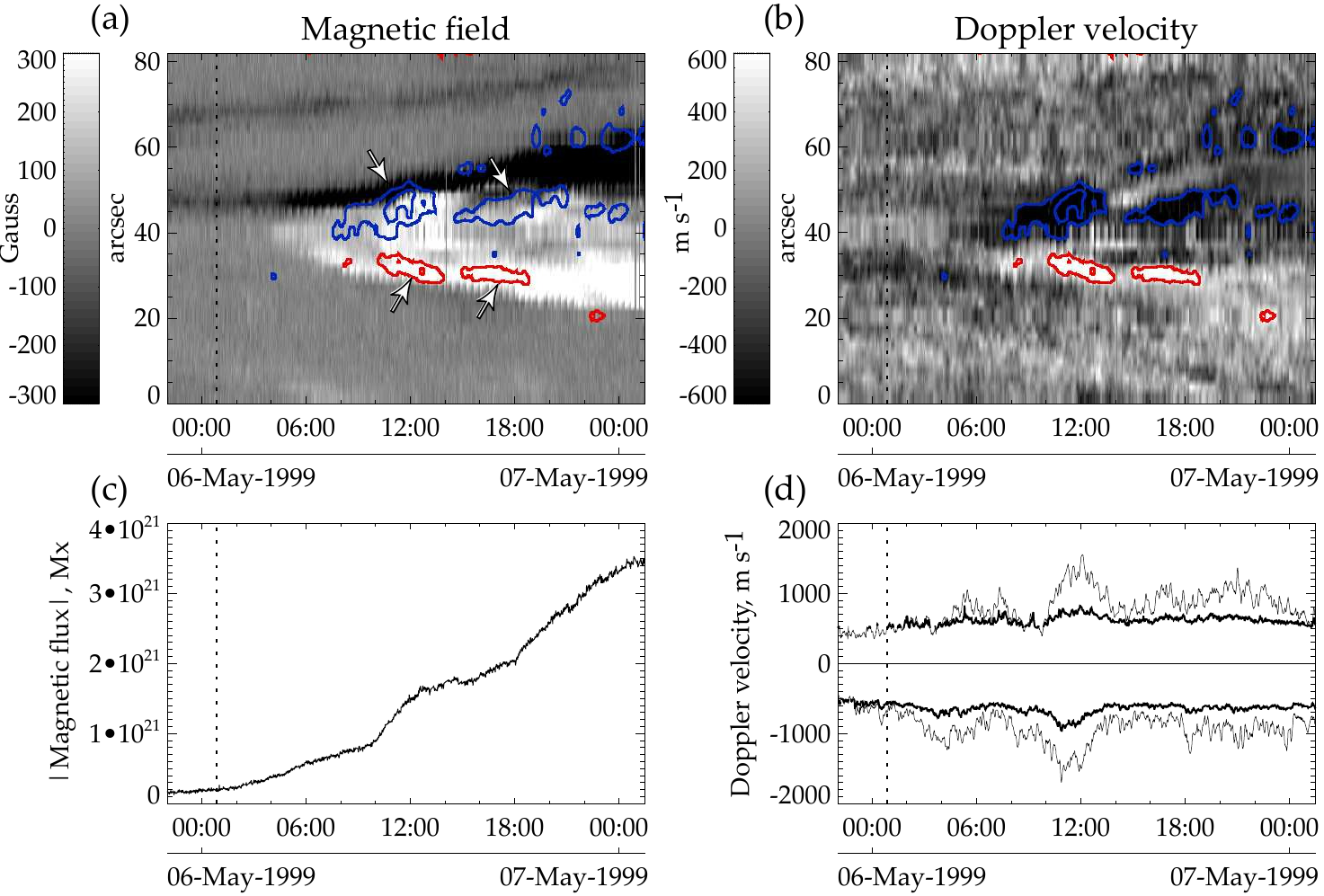}
}
\caption{
Active region NOAA 8536: time slice diagrams of (a) magnetic field strength and
(b) Doppler velocities. The location of the slice is marked on the magnetograms
in Figure~\ref{fig:02} by a black line. The marking convention is the same as
in Figure~\ref{fig:05}.
	}
\label{fig:06}
\end{figure}

\begin{figure}
\centerline{
\includegraphics[width=11.53cm]{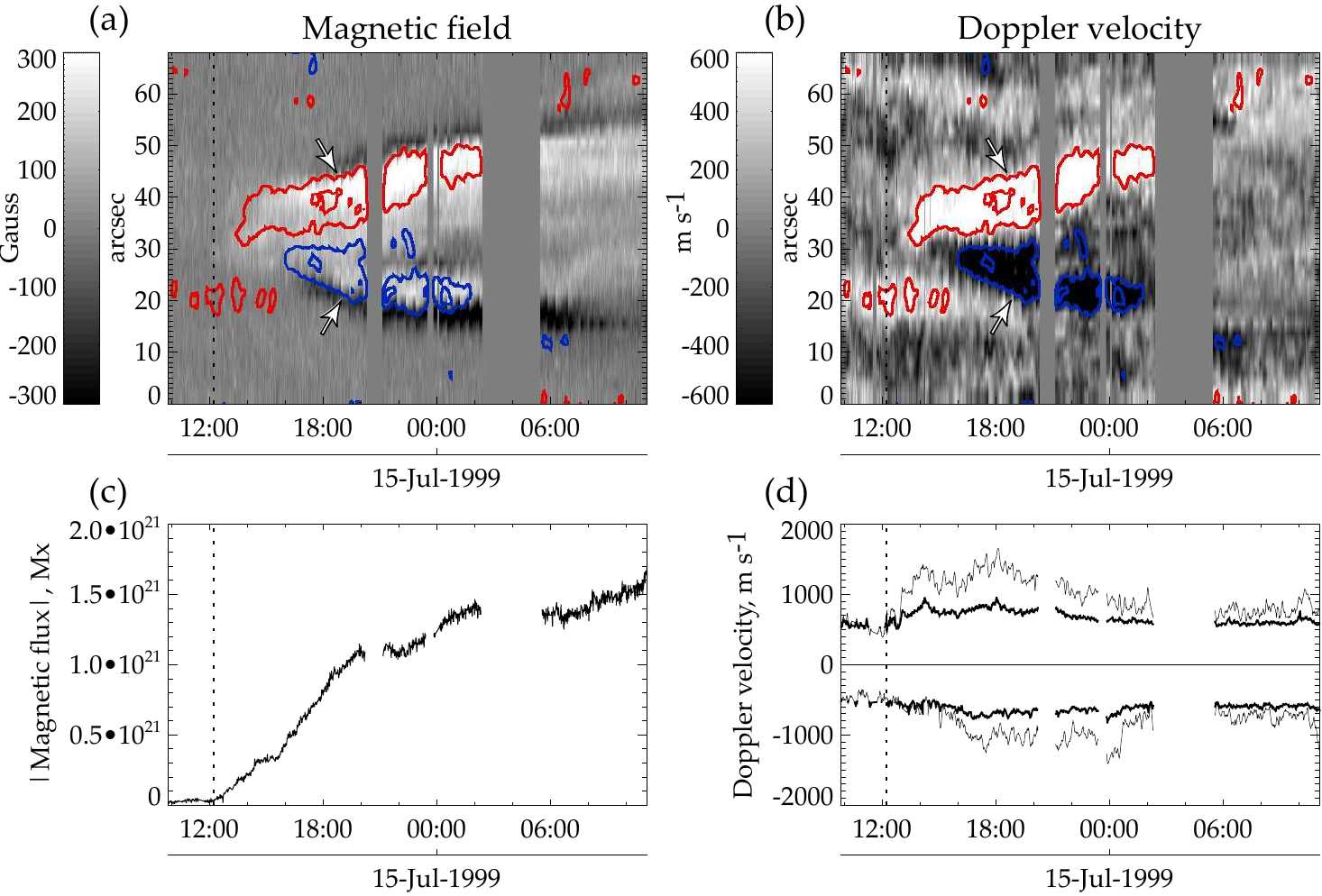}
}
\caption{
Active region NOAA 8635: time slice diagrams of (a) magnetic field strength and
(b) Doppler velocities. The location of the slice is marked on the magnetograms
in Figure~\ref{fig:03} by a black line. The marking convention is the same as
in Figure~\ref{fig:05}.
	}
\label{fig:07}
\end{figure}

\begin{figure}
\centerline{
\includegraphics[width=11.53cm]{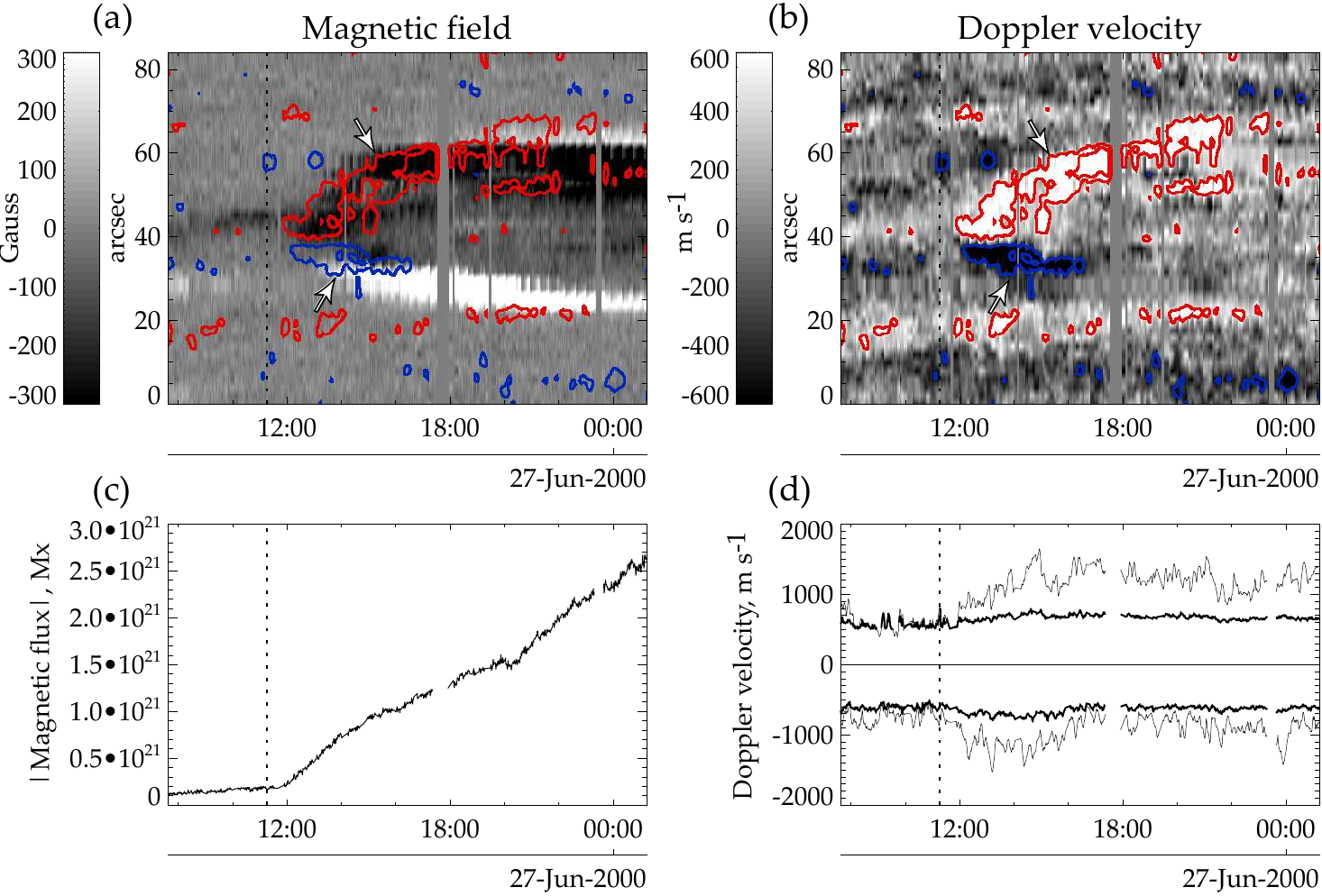}
}
\caption{
Active region NOAA 9064: time slice diagrams of (a) magnetic field strength and
(b) Doppler velocities. The location of the slice is marked on the magnetograms
in Figure~\ref{fig:04} by a black line. The marking convention is the same as
in Figure~\ref{fig:05}.
	}
\label{fig:08}
\end{figure}

 Time slice diagrams for magnetic field strength and Doppler velocities were
constructed in the first hours of emergence of active regions (panels (a) and
(b) in Figures~\ref{fig:05}, \ref{fig:06}, \ref{fig:07}, and \ref{fig:08}). The
slices are orientated along the axis of the emerging bipolar pairs, and their
position is marked by a black line on the magnetograms in Figures~\ref{fig:01},
\ref{fig:02}, \ref{fig:03}, and \ref{fig:04}. During the emergence of active
regions a slight rotation of the axis of dipoles is observed; therefore, at the
very beginning of magnetic flux emergence both polarities with Doppler velocity
structures do not always fall into the slices.

\par The time slice diagrams clearly show that the Doppler
velocity structures are located within the limits of emerging magnetic
flux, and that they are not present in the surrounding regions
(Figures~\ref{fig:05}\,a, \ref{fig:06}\,a, \ref{fig:07}\,a, \ref{fig:08}\,a).
The Doppler velocity structures form at the beginning of magnetic flux
emergence, occupy an extensive region, and persist for a few hours
(Figures~\ref{fig:05}\,b, \ref{fig:06}\,b, \ref{fig:07}\,b, \ref{fig:08}\,b and
Table~\ref{tbl:dop-vel}).  Regions of Doppler velocities of different signs do
not appear at the same time. In the first hours of magnetic flux
emergence they are adjacent to each other; they then separate together with the
opposite polarities.

\par The Doppler velocity values increase gradually and reach their peak values
4\,--\,12 hours after the start of the emergence of active regions,
approximately at half of the velocity structure's life time
(Figures~\ref{fig:05}\,d, \ref{fig:06}\,d, \ref{fig:07}\,d, \ref{fig:08}\,d).
The peak values of mean Doppler velocity inside the $\pm$500\,{\ms} isolines are
800\,--\,970\,{\ms}, and the peak values of maximum Doppler velocities reach
1410\,--\,1700\,{\ms} (Table~\ref{tbl:dop-vel}). The mean Doppler velocities
show the presence of velocity structures rather weakly. At the same time, as
noted above, the Doppler velocities inside the $\pm$1000\,{\ms} isolines, within
individual time intervals, occupy significant areas which exist for a long time.
The peak values of the Doppler velocities accompanying the emergence of active
regions (Table~\ref{tbl:dop-vel}) substantially exceed the maximum Doppler
velocities of the convective flow of the quiet Sun, which, in the SOHO/MDI data,
do not exceed 1200\,{\ms} (the points with $\theta>50^{\circ}$ in Figure 3\,b of
\opencite{khl11}).

\par We calculated the separation velocities of the photospheric magnetic flux
outer boundaries $V_{sep}$, and estimated the contribution of magnetic flux
expansion to the Doppler velocity signal $V_{exp}$ (Table~\ref{tbl:sep-vel}).
From the comparison in Table~\ref{tbl:dop-vel} and Table~\ref{tbl:sep-vel} one
can see that the Doppler velocity values observed substantially exceed the
magnetic flux expansion velocity $V_{exp}$.

\par Previews of other active regions emerging near the limb showed that these
powerful and long-lived Doppler velocity structures do not always appear. They
seem to be present only in events with a high flux growth rate.
\inlinecite{khl12} has carried out a statistical investigation of 54 active
regions with different spatial scales (total unsigned magnetic flux
$8\times10^{19}$\,--\,$5\times10^{22}$\,Mx) emerging near the limb. It was found
that the peak values of negative and positive Doppler velocities are
related quadratically to the magnetic flux growth rate in the first hours of the
emergence of active regions (Figure 6a of \opencite{khl12}).


\begin{table}
\begin{tabular}{cccccccccc}
\hline
{\bf Active}&								
\multicolumn{4}{c}{\bf Negative Doppler velocities}&		
\multicolumn{4}{c}{\bf Positive Doppler velocities}		
\\
{\bf regions}&								
{$\bf V_{mean-},$}&							
{$\bf V_{max-},$}&							
{\bf d,} \mynote{Size of Doppler velocity structure inside the isoline $-$500 or $+$500\,{\ms} along the slices \\ in Figures~\ref{fig:05}\,b, \ref{fig:06}\,b, \ref{fig:07}\,b, \ref{fig:08}\,b taking into account the projection effect.}&
{\bf t,}&								
{$\bf V_{mean+},$}&							
{$\bf V_{max+},$}&							
{\bf d,} \mynotemark[1]&						
{\bf t,}								
\\
{\bf }&									
{$\bf m\,s^{-1}$}&							
{$\bf m\,s^{-1}$}&							
{\bf Mm}&								
{\bf h}&								
{$\bf m\,s^{-1}$}&							
{$\bf m\,s^{-1}$}&							
{\bf Mm}&								
{\bf h}&								
\\ \hline
9037 & $-$810 & $-$1640 & 14  & 15   & 850 & 1630 & 15 & 13   \\
8536 & $-$970 & $-$1700 & 18  & 14   & 830 & 1560 & 10 &  9   \\
8635 & $-$830 & $-$1410 & 16.5& 10   & 960 & 1660 & 15 & 13.5 \\
9064 & $-$800 & $-$1520 &  9  &  4.5 & 810 & 1650 & 16 & 10
\\ \hline
\end{tabular}
\themynotes
\caption[]{Peak values of mean and absolut maximum Doppler velocities
($V_{mean-}$, $V_{max-}$, $V_{mean+}$, $V_{max+}$), size ($d$), and life time
($t$) of the velocity structures during the first hours of the emergence of
active regions}
\label{tbl:dop-vel}
\end{table}

\begin{table}
\begin{tabular}{cccc}
\hline
{\bf Active}&								
{\bf Time interval}&  							
{$\bf \overrightarrow{V}_{sep},$}&					
{$\bf V_{exp},$}							
\\
{\bf regions}&								
{\bf }&									
{$\bf m\,s^{-1}$}&							
{$\bf m\,s^{-1}$}							
\\
\hline
9037 & 10 Jun 2000, 17:10 -- 19:10 UT & 450 & $\lesssim$ 370 \\
8536 & 06 May 1999, 10:05 -- 12:05 UT & 150 & $\lesssim$ 130 \\
8635 & 14 Jul 1999, 17:00 -- 19:00 UT & 340 & $\lesssim$ 290 \\
9064 & 26 Jun 2000, 14:00 -- 16:00 UT & 820 & $\lesssim$ 660
\\ \hline
\end{tabular}
\caption[]{The mean relative separation velocities of the photospheric magnetic flux outer boundaries ($\overrightarrow{V}_{sep}$), the contribution of the magnetic flux expansion to the line of sight signal of Doppler velocity ($V_{exp}$), and the periods of time (Time intervals) for which $\overrightarrow{V}_{sep}$ and $V_{exp}$ were determined}
\label{tbl:sep-vel}
\end{table}

\section{Geometrical Asymmetry of Emerging Magnetic Flux}

\par During the appearance of active regions it is possible to see the
well-known geometrical asymmetry of emerging magnetic flux (\opencite{dri90} and
references therein). It appears that sunspots of leading polarity move away from the
location of the emergence much faster than those of following polarity do. In
the time slice diagrams of the magnetic field of the active regions under
consideration, the following polarity is situated at the bottom, and the leading
one at the top (Figures~\ref{fig:05}\,a, \ref{fig:06}\,a, \ref{fig:07}\,a,
\ref{fig:08}\,a). In NOAA 9037, soon after emergence, the following (negative)
polarity is set against the existing concentration of the positive magnetic
field and stops drifting, while the leading polarity moves along the slice at
high velocity (Figure~\ref{fig:05}\,a). NOAA 8536 emerges before an existing
concentration of negative magnetic field. Therefore, the following polarity
moves faster than the leading one in the first hours of emergence, but then the
leading polarity begins to move with greater velocity (Figure~\ref{fig:06}\,a).
In NOAA 8635 and 9064, emerging at the W-limb, it is clearly seen that the
leading polarity moves considerably faster than the following one
(Figures~\ref{fig:07}\,a and \ref{fig:08}\,a).

\begin{figure}
\centerline{
\includegraphics[width=\textwidth]{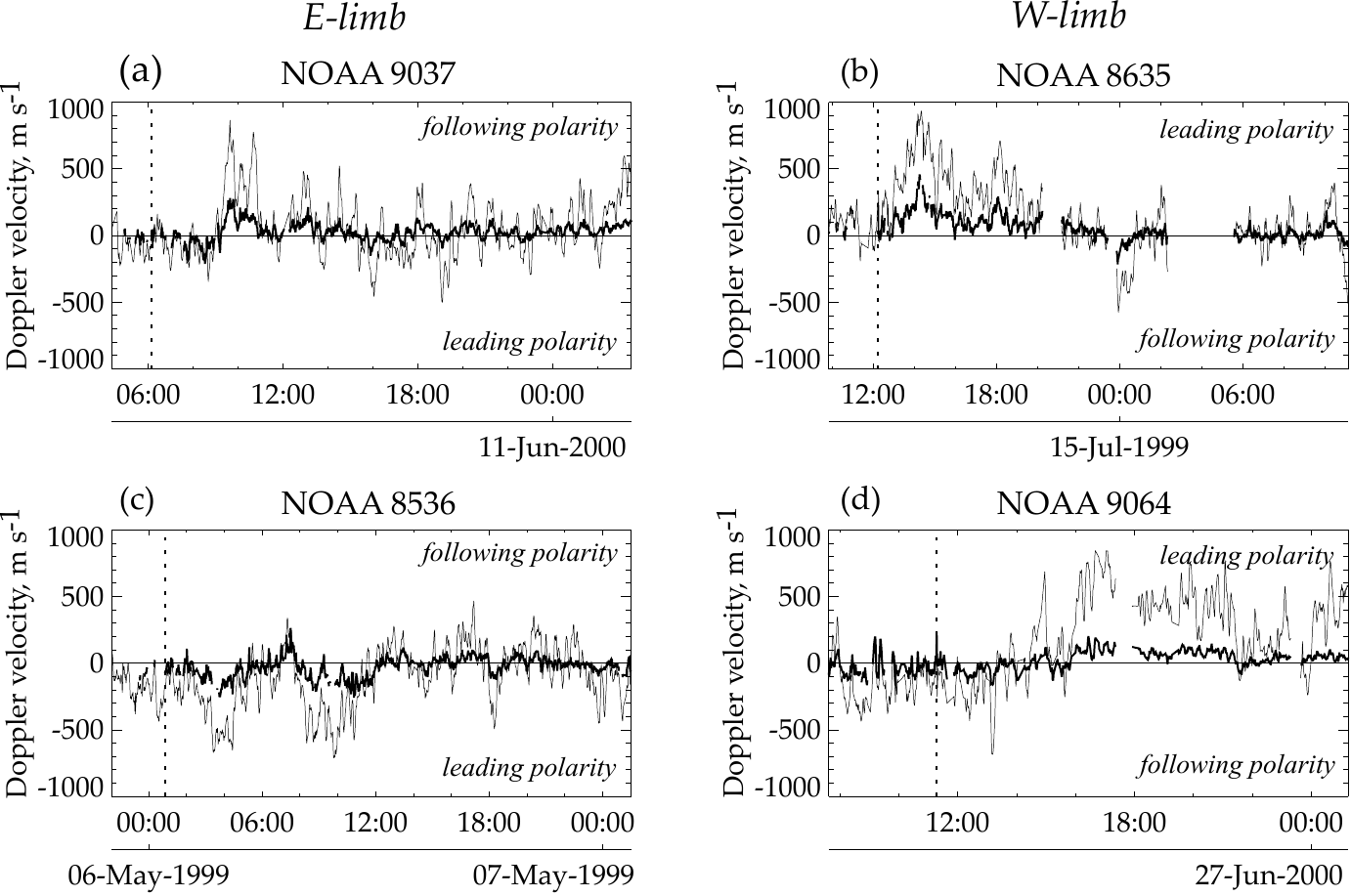}
}
\caption{
The time variation for the difference between the positive and negative Doppler
velocities for mean (thick line) and maximal (thin line) values in the studied
active regions. The polarities of the locations of the positive and negative
Doppler velocities are marked on the plots.
	}
\label{fig:09}
\end{figure}

\section{Asymmetry in the Doppler Velocity Fields}

\par Asymmetry is also observed in the Doppler velocity fields.
In the time slice diagrams in Figures~\ref{fig:05}\,a, \ref{fig:06}\,a,
\ref{fig:07}\,a, \ref{fig:08}\,a, one can clearly see that the Doppler velocity
structures located at the region of the leading magnetic polarity are more
powerful and exist longer than in the following one.

\par We performed a quantitative analysis for the Doppler velocity asymmetry.
Figures~\ref{fig:09} presents the plots of time variation for the difference
between the positive and negative Doppler velocities for mean and maximal
values. There are some blanks in the mean Doppler velocity asymmetry for those
time instants when the velocities in the region of emerging magnetic
flux were less than 500\,{\ms}. For active regions NOAA 8536 and 8635
emerging near different limbs, a well-defined dominance of the Doppler
velocities located in the leading polarity is observed during the first
12 hours of the emergence. In two other active regions, NOAA 9037 and NOAA 9064,
also emerging near different limbs, the asymmetry value changes its sign. The
NOAA 9037 appearance starts with the emergence of a magnetic loop whose axis
is oriented almost perpendicularly to the line of sight. Its rise is accompanied
by a dominance of the negative Doppler velocity. Approximately from 09:00 UT,
a magnetic flux emergence appears whose axis is oriented along the
line of sight. Against this background, one observes a significant dominance of
the positive Doppler velocities localized in the following polarity. In the NOAA
9064 early emergence, the mean Doppler velocity values show practically no
asymmetry, but one observes a dominance of the maximal negative Doppler
velocities corresponding to the following polarity. Approximately three hours after
the start of emergence, the asymmetry value changes sign, and a noticeable
dominance of the positive Doppler velocity localized in the leading polarity
begins.

\par In all the active regions under consideration, the Doppler velocity
asymmetry reaches its peak value soon after the start of magnetic flux
emergence, and then gradually drops within 7\,--\,12 hours. The Doppler
velocity asymmetry decreasing time is comparable to the lifetime of
the velocity structures. The peak values of asymmetry for the mean and
maximal Doppler velocities reach 240\,--\,460\,{\ms} and 710\,--\,940\,{\ms},
respectively.

\section{Interpretation of the Results}

\par
\begin{figure}
\centerline{
\includegraphics[width=12cm]{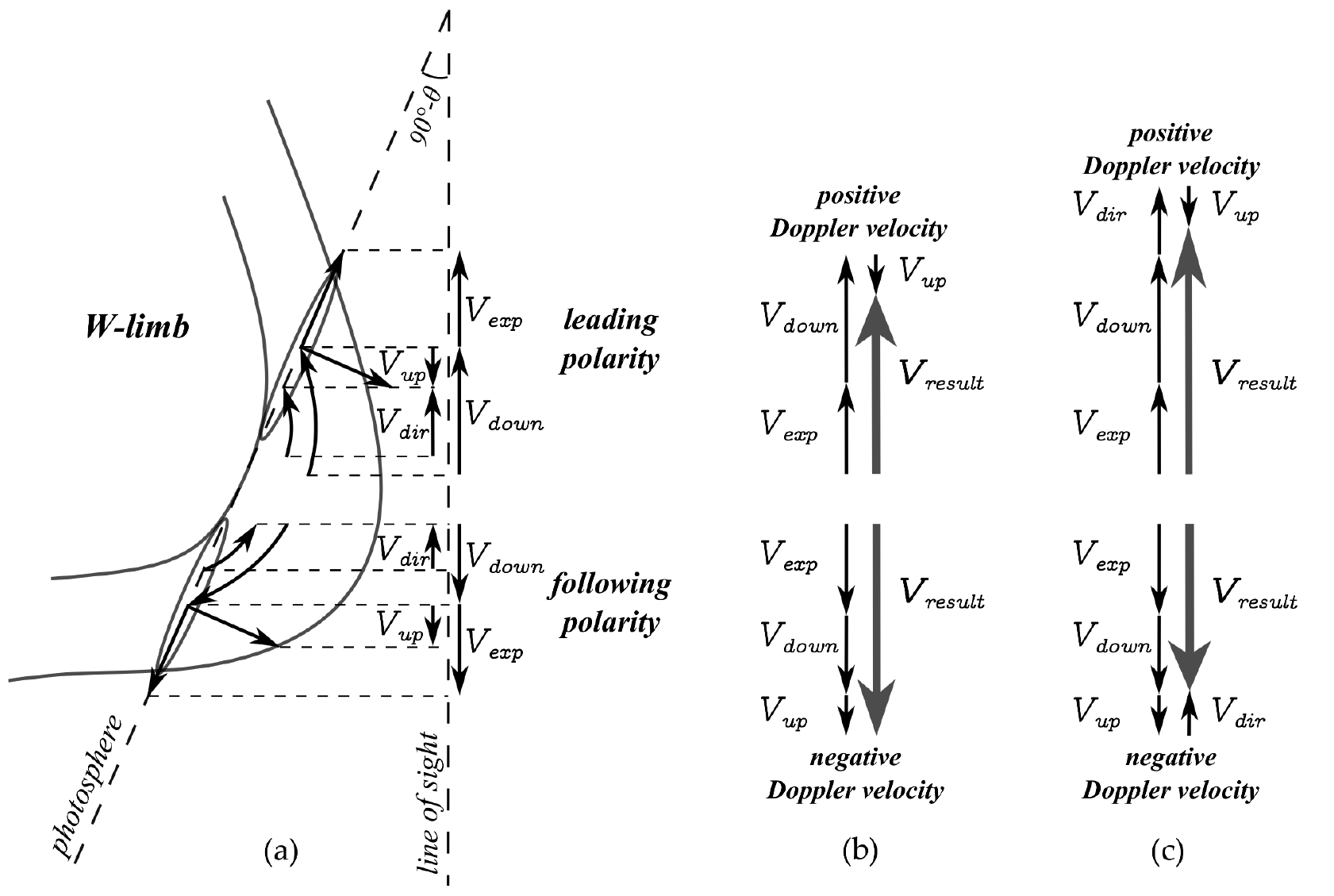}
}
\caption{
(a) The simple scheme of the plasma flows accompanying the emergence of
the magnetic flux near the W-limb. The velocity vectors of the possible
flows and their projection to the line of sight in the leading and following
polarities are marked by arrows. $V_{up}$ is the magnetic flux
rise velocity; $V_{down}$ is the plasma downflow velocity being carried
out into the solar atmosphere by the emerging magnetic flux; $V_{exp}$
is the magnetic flux horizontal expansion velocity; $V_{dir}$ is the plasma
directional flow velocity inside the emerging magnetic structure (in this case
from the following polarity to the leading one). (b) The compositional
result of the velocity components along the line of sight without the plasma
directional flow $V_{dir}$. (c) The compositional result of the velocity
components along the line of sight taking into account the plasma directional
flow $V_{dir}$.
	}
\label{fig:10}
\end{figure}

 Let us consider the contribution of possible motions to the line of sight
Doppler velocity signal in data with low spatial resolution (Figure~\ref{fig:10}\,a).

\par \textit{i)} \textit{The magnetic flux rise velocity
$\overrightarrow{V}_{up}$.} $\overrightarrow{V}_{up}$ should have a similar
contribution in the leading and following polarities. When active regions emerge
near the solar disk center, the maximum values of the upflow Doppler velocities
reach $\overrightarrow{V}_{up}\sim$300\,--\,1000\,{\ms} (Figure 4 of
\opencite{khl12} and references in Section 1 of this paper). The projection of
these velocities to the line of sight at the heliocentric angle
$\theta$\,=\,$60^\circ$ will be $V_{up} \lesssim$150\,--\,500\,{\ms}.

\par \textit{ii)} \textit{The plasma downflow velocity
$\overrightarrow{V}_{down}$.} In the active regions emerging near the solar disk
center, at the footpoints of the magnetic loops, one observes positive Doppler
velocities interpreted as a downflow of plasma being carried out into the solar
atmosphere (see references in Section 1 of this paper). The plasma downflow
takes place along the magnetic field lines. Theoretical models show that the
velocities of draining plasma have a significant horizontal component at the
beginning of the magnetic flux emergence ({\it e.g.,} \opencite{shi90},
\opencite{arch04}, \opencite{tor10}, \opencite{tor11}).

\par\textit{iii)} \textit{The magnetic flux horizontal expansion velocity
$\overrightarrow{V}_{exp}$.} $\overrightarrow{V}_{exp}$ should have a similar
contribution in the leading and the following polarities under the condition of
equality of their gas pressure. From the theory of magnetic flux emergence
into the solar atmosphere, an inverse relation exists between the expansion
velocities and degree of twist of the emerging magnetic flux (see, for example,
\opencite{mur06}). \inlinecite{che10} showed that an emerging tube with a total
toroidal flux content of $7.6\times10^{21}$\,Mx creates a transient pressure
excess that results in diverging horizontal flows with velocities over 3\,{\kms}
for approximately five hours. In this case, the plasma flow velocities are
comparable to the separation velocities of the photospheric magnetic flux outer
boundaries. The observations show that at the very beginning of the emergence
the magnetic flux expansion has a maximal velocity that then drops rapidly with
time. Even at the emergence of the powerful active region NOAA 10488, with a
total unsigned magnetic flux at the maximum evolution of $>6\times10^{22}$\,Mx and a
total unsigned magnetic flux growth rate during the first hours of
$4.1\times10^{20}$\,Mx\,h$^{-1}$, the separation velocities of the photospheric
magnetic flux outer boundaries had already dropped to 0.3\,{\kms} two hours
after the beginning of the emergence \cite{gri09}.

\par \textit{iv)} \textit{It is possible that plasma directional flows exist
inside the emerging magnetic structure $\overrightarrow{V}_{dir}$.} Directional
flows, as well as the plasma downflow $\overrightarrow{V}_{down}$, take place
along the magnetic field lines. Searches for such flows in the Doppler velocity
asymmetry between the leading and following polarities in the active regions
emerging near the solar disk center did not provide consistent results
\cite{cau96,sig98,cha02,pev06,bat06,gri11}. Plasma flows are theoretically
expected to be directed from the leading polarity into the following one. The
thin flux tube theory shows that these flows are due to the Coriolis force
acting on the emerging magnetic structure, and the velocities of these flows
reach some hundreds of meters per second (for a detailed analysis, see
\opencite{fan09}). However, when entering the convection zone high layers
(20\,--\,30\,Mm under the photosphere), the magnetic flux undergoes a strong
expansion and fragmentation. Therefore, the probability of conservation of the
flows caused by the Coriolis force is not known. The models of the magnetic flux
emergence from the near-surface layers into the solar atmosphere do not consider
the existence of these flows. It is also possible that plasma flows occur from
the following polarity into the leading one. A discussion of the mechanisms
leading to their origin may be found in \inlinecite{cau96}.

\par Thus, the line of sight Doppler velocity signal in data with low
spatial resolution contains contributions from the following flows
(Figure~\ref{fig:10}\,a):
\begin{eqnarray}
\overrightarrow{V} = \overrightarrow{V}_{up} + \overrightarrow{V}_{down} + \overrightarrow{V}_{exp} + \overrightarrow{V}_{dir},
\end{eqnarray}
Their projections onto the line of sight in the positive and negative
Doppler velocity structures are determined as:
\begin{eqnarray}
V_{-} = +\overrightarrow{V}_{up} cos \theta + \overrightarrow{V}_{down} cos X_{1} + \overrightarrow{V}_{exp} sin \theta \pm \overrightarrow{V}_{dir} cos X_{1},
\end{eqnarray}
\begin{eqnarray}
V_{+} = -\overrightarrow{V}_{up} cos \theta + \overrightarrow{V}_{down} cos X_{2} + \overrightarrow{V}_{exp} sin \theta \pm \overrightarrow{V}_{dir} cos X_{2},
\end{eqnarray}
where $\theta$ is the heliocentric angle, and $X_{1}$ and $X_{2}$ are the
angles between the line of sight and the magnetic field lines in the polarities
where the negative and positive velocity structures are localized.

\par In the velocity structures accompanying the emergence of active regions
under consideration, one observes high Doppler velocities that reach their peak
values 4\,--\,12 hours after the magnetic flux emergence begins
(Figures~\ref{fig:05}\,d, \ref{fig:06}\,d, \ref{fig:07}\,d, \ref{fig:08}\,d).
Doppler velocities more than 1000\,{\ms} concentrate in the central part of the
velocity structures, occupy significant areas, and exist over long time
intervals (Figures~\ref{fig:01}, \ref{fig:02}, \ref{fig:03}, and \ref{fig:04}).
Let us discuss what the observed velocities are associated with, having
considered the contribution of possible motions to the line of sight Doppler
velocity signal at the heliocentric angle $\theta=60^\circ$. The maximal
contribution of the magnetic flux rise velocities should be
$V_{up}$\,$\sim$\,150\,--\,500\,{\ms}, with a mean contribution of 300\,{\ms}.
The contribution of the magnetic flux horizontal expansion velocities to our
active regions is $V_{exp}$\,$\sim$\,130\,--\,660\,{\ms}
(Table~\ref{tbl:sep-vel}). The velocity of directional flow
$\overrightarrow{V}_{dir}$ will determine the value of the Doppler velocity
asymmetry between the velocity structures of opposite sign and, for active
regions NOAA 9037 and 8536, 12 hours after the start of emergence it has no
contribution. Thus, one can see that the high Doppler velocities are caused by
the significant component of the plasma downflow velocities
$\overrightarrow{V}_{down}$ (Figure~\ref{fig:10}\,b).

\par The Doppler velocity asymmetry between velocity structures of the leading
and following polarities reaches its peak values soon after the emergence
begins, and then it gradually drops (Figure~\ref{fig:09}). In two active
regions, NOAA 8536 and 8635, emerging near different limbs, one observes an
explicit dominance of the Doppler velocities localized in the leading polarity.
In two other active regions, NOAA 9037 and 9064, also emerging near different
limbs, the Doppler velocity asymmetry value changes sign, but, at individual
time intervals, there is a Doppler velocity dominance in the following polarity.
There is probably a contribution from the plasma directional flow $V_{dir}$
(Figure~\ref{fig:10}\,c). The Doppler velocity asymmetry may be caused by a
morphological asymmetry of active regions that becomes apparent because the
magnetic flux of the leading polarity is more compact than of the following one.
However, to better understand the causes for the asymmetry between Doppler
velocity structures, one should search for some additional relations.

\section{Conclusion} 
  \label{S-Conclusion}

\par Analysis of photospheric flows accompanying the emergence of active regions
near the limb showed the existence of an extensive region of enhanced negative
Doppler velocities on the boundary where the magnetic field changes sign, which
is located in the polarity closer to the solar disk center because of the projection effect.
The region of positive Doppler velocities is adjacent to it in the magnetic
polarity closer to the solar limb. The observed flows form at the start of
active region emergence and are present for a few hours. The Doppler velocity
values increase gradually and reach their peak values 4\,--\,12 hours after the
start of the magnetic flux emergence. The peak values of the mean (inside the
$\pm$500\,{\ms} isolines) and maximum Doppler velocities are 800\,--\,970\,{\ms}
and 1410\,--\,1700\,{\ms}, respectively. The Doppler velocity values observed
substantially exceed the separation velocities of the photospheric magnetic flux
outer boundaries and most likely are caused by a significant component of the
plasma downflow velocities being carried out into the solar atmosphere by emerging
magnetic flux.

\par An asymmetry was detected between the velocity structures of the leading
and following polarities. Doppler velocity structures located in a region of
leading magnetic polarity are more powerful and exist longer than those in a
region of following magnetic polarity. The Doppler velocity
asymmetry between velocity structures of the leading and following polarities
reaches peak values soon after emergence begins and then gradually drops within
7\,--\,12 hours. The peak values of asymmetry for the mean and maximal Doppler
velocities reach 240\,--\,460\,{\ms} and 710\,--\,940\,{\ms}, respectively.
Additional investigations are necessary to understand the reasons for the
Doppler velocity asymmetry. However, it could be caused by the plasma
directional flows inside the emerging magnetic structure or morphological
asymmetry between the leading and following polarities.

\begin{acks}
The author expresses sincere gratitude to the Geuest Editor for useful comments.
The author is grateful to Prof. V.M. Grigoriev, L.V. Ermakova, and V.G.
Eselevich for important suggestions and help in understanding the obtained
results. This work used data obtained by the SOHO/MDI instrument. SOHO is a
mission of international cooperation between ESA and NASA. The MDI is a project
of the Stanford-Lockheed Institute for Space Research. This study was supported
by RFBR grants 10-02-00607-a, 10-02-00960-a, 11-02-00333-a, 12-02-00170-a, state
contracts of the Ministry of Education and Science of the Russian Federation
Nos. 02.740.11.0576 and 16.518.11.7065, the program of the Division of Physical
Sciences of the Russian Academy of Sciences No. 16, the Integration Project of
SB RAS No. 13, and the program of the Presidium of Russian Academy of Sciences
No. 22.
\end{acks}

\bibliographystyle{spr-mp-sola}
\bibliography{khlystova2}

\begin{thebibliography}{47}
\ifx \bisbn   \undefined \def \bisbn  #1{ISBN #1}\fi
\ifx \binits  \undefined \def \binits#1{#1}\fi
\ifx \bauthor  \undefined \def \bauthor#1{#1}\fi
\ifx \batitle  \undefined \def \batitle#1{#1}\fi
\ifx \bjtitle  \undefined \def \bjtitle#1{\textit{#1}}\fi
\ifx \bvolume  \undefined \def \bvolume#1{\textbf{#1}}\fi
\ifx \byear  \undefined \def \byear#1{#1}\fi
\ifx \bissue  \undefined \def \bissue#1{#1}\fi
\ifx \bfpage  \undefined \def \bfpage#1{#1}\fi
\ifx \blpage  \undefined \def \blpage #1{#1}\fi
\ifx \burl  \undefined \def \burl#1{\textsf{#1}}\fi
\ifx \href  \undefined \def \href#1#2{\textsf{#2}}\fi
\ifx \doiurl  \undefined \def
  \doiurl#1{\href{http://dx.doi.org/#1}{\textsf{#1}}}\fi
\ifx \betal  \undefined \def \betal{\textit{et al.}}\fi
\ifx \binstitute  \undefined \def \binstitute#1{#1}\fi
\ifx \bctitle  \undefined \def \bctitle#1{#1}\fi
\ifx \beditor  \undefined \def \beditor#1{#1}\fi
\ifx \bpublisher  \undefined \def \bpublisher#1{#1}\fi
\ifx \bbtitle  \undefined \def \bbtitle#1{\textit{#1}}\fi
\ifx \bedition  \undefined \def \bedition#1{#1}\fi
\ifx \bseriesno  \undefined \def \bseriesno#1{\textbf{#1}}\fi
\ifx \blocation  \undefined \def \blocation#1{#1}\fi
\ifx \bsertitle  \undefined \def \bsertitle#1{\textit{#1}}\fi
\ifx \bsnm \undefined \def \bsnm#1{#1}\fi
\ifx \bsuffix \undefined \def \bsuffix#1{#1}\fi
\ifx \bparticle \undefined \def \bparticle#1{#1}\fi
\ifx \barticle \undefined \def \barticle#1{}\fi
\ifx \botherref \undefined \def \botherref#1{}\fi
\ifx \url \undefined \def \url#1{\textsf{#1}}\fi
\ifx \bchapter \undefined \def \bchapter#1{}\fi
\ifx \bbook \undefined \def \bbook#1{}\fi
\ifx \bcomment \undefined \def \bcomment#1{#1}\fi
\ifx \oauthor \undefined \def \oauthor#1{#1}\fi
\ifx \citeauthoryear \undefined \def \citeauthoryear#1{#1}\fi
\def \endbibitem {}
\ifx \bconflocation  \undefined \def \bconflocation#1{#1} \fi

\bibitem[\protect\citeauthoryear{{Archontis} \textit{et~al.}}{2004}]{arch04}
\begin{barticle}
\bauthor{\bsnm{{Archontis}}, \binits{V.}},
\bauthor{\bsnm{{Moreno-Insertis}}, \binits{F.}},
\bauthor{\bsnm{{Galsgaard}}, \binits{K.}},
\bauthor{\bsnm{{Hood}}, \binits{A.}},
\bauthor{\bsnm{{O'Shea}}, \binits{E.}}:
\byear{2004},
\batitle{{Emergence of magnetic flux from the convection zone into the
  corona}}.
\bjtitle{\aap}
\bvolume{426},
\bfpage{1047}\,--\,\blpage{1063}.
doi:\doiurl{10.1051/0004-6361:20035934}.
\end{barticle}
\endbibitem

\bibitem[\protect\citeauthoryear{{Bachmann}}{1978}]{bac78}
\begin{barticle}
\bauthor{\bsnm{{Bachmann}}, \binits{G.}}:
\byear{1978},
\batitle{{On the evolution of magnetic and velocity fields of an originating
  sunspot group}}.
\bjtitle{Bulletin of the Astronomical Institutes of Czechoslovakia}
\bvolume{29},
\bfpage{180}\,--\,\blpage{184}.
\end{barticle}
\endbibitem

\bibitem[\protect\citeauthoryear{{Barth} and {Livi}}{1990}]{bar90}
\begin{barticle}
\bauthor{\bsnm{{Barth}}, \binits{C.S.}},
\bauthor{\bsnm{{Livi}}, \binits{S.H.B.}}:
\byear{1990},
\batitle{{Magnetic Bipoles in Emerging Flux Regions on the Sun}}.
\bjtitle{Rev. Mex. Astron. Astrofis.}
\bvolume{21},
\bfpage{549}.
\end{barticle}
\endbibitem

\bibitem[\protect\citeauthoryear{{Battiato} \textit{et~al.}}{2006}]{bat06}
\begin{bchapter}
\bauthor{\bsnm{{Battiato}}, \binits{V.}},
\bauthor{\bsnm{{Billotta}}, \binits{S.}},
\bauthor{\bsnm{{Contarino}}, \binits{L.}},
\bauthor{\bsnm{{Guglielmino}}, \binits{S.}},
\bauthor{\bsnm{{Romano}}, \binits{P.}},
\bauthor{\bsnm{{Soadaro}}, \binits{D.}},
\bauthor{\bsnm{{Zuccarello}}, \binits{F.}}:
\byear{2006},
\bctitle{{High Resolution Observations of Emerging Active Regions Carried Out
  at the THEMIS Telescope}}.
In: \bbtitle{SOHO-17. 10 Years of SOHO and Beyond},
\bsertitle{ESA Special Publication}
\bseriesno{617}.
\end{bchapter}
\endbibitem

\bibitem[\protect\citeauthoryear{{Brants}}{1985a}]{bra85a}
\begin{barticle}
\bauthor{\bsnm{{Brants}}, \binits{J.J.}}:
\byear{1985}a,
\batitle{{High-resolution spectroscopy of active regions. II Line-profile
  interpretation, applied to an emerging flux region}}.
\bjtitle{\solphys}
\bvolume{95},
\bfpage{15}\,--\,\blpage{36}.
doi:\doiurl{10.1007/BF00162633}.
\end{barticle}
\endbibitem

\bibitem[\protect\citeauthoryear{{Brants}}{1985b}]{bra85b}
\begin{barticle}
\bauthor{\bsnm{{Brants}}, \binits{J.J.}}:
\byear{1985}b,
\batitle{{High-resolution spectroscopy of active regions. III - Relations
  between the intensity, velocity, and magnetic structure in an emerging flux
  region}}.
\bjtitle{\solphys}
\bvolume{98},
\bfpage{197}\,--\,\blpage{217}.
doi:\doiurl{10.1007/BF00152456}.
\end{barticle}
\endbibitem

\bibitem[\protect\citeauthoryear{{Brants} and {Steenbeek}}{1985}]{bra85c}
\begin{barticle}
\bauthor{\bsnm{{Brants}}, \binits{J.J.}},
\bauthor{\bsnm{{Steenbeek}}, \binits{J.C.M.}}:
\byear{1985},
\batitle{{Morphological evolution of an emerging flux region}}.
\bjtitle{\solphys}
\bvolume{96},
\bfpage{229}\,--\,\blpage{252}.
doi:\doiurl{10.1007/BF00149682}.
\end{barticle}
\endbibitem

\bibitem[\protect\citeauthoryear{{Cauzzi}, {Canfield}, and
  {Fisher}}{1996}]{cau96}
\begin{barticle}
\bauthor{\bsnm{{Cauzzi}}, \binits{G.}},
\bauthor{\bsnm{{Canfield}}, \binits{R.C.}},
\bauthor{\bsnm{{Fisher}}, \binits{G.H.}}:
\byear{1996},
\batitle{{A Search for Asymmetric Flows in Young Active Regions}}.
\bjtitle{\apj}
\bvolume{456},
\bfpage{850}\,--\,\blpage{860}.
doi:\doiurl{10.1086/176702}.
\end{barticle}
\endbibitem

\bibitem[\protect\citeauthoryear{{Chapman}}{2002}]{cha02}
\begin{barticle}
\bauthor{\bsnm{{Chapman}}, \binits{G.A.}}:
\byear{2002},
\batitle{{A Study of AR 9144; A Fast-Growing EFR}}.
\bjtitle{\solphys}
\bvolume{209},
\bfpage{141}\,--\,\blpage{152}.
doi:\doiurl{10.1023/A:1020994131849}.
\end{barticle}
\endbibitem

\bibitem[\protect\citeauthoryear{{Cheung} \textit{et~al.}}{2010}]{che10}
\begin{barticle}
\bauthor{\bsnm{{Cheung}}, \binits{M.C.M.}},
\bauthor{\bsnm{{Rempel}}, \binits{M.}},
\bauthor{\bsnm{{Title}}, \binits{A.M.}},
\bauthor{\bsnm{{Sch{\"u}ssler}}, \binits{M.}}:
\byear{2010},
\batitle{{Simulation of the Formation of a Solar Active Region}}.
\bjtitle{\apj}
\bvolume{720},
\bfpage{233}\,--\,\blpage{244}.
doi:\doiurl{10.1088/0004-637X/720/1/233}.
\end{barticle}
\endbibitem

\bibitem[\protect\citeauthoryear{{Chou} and {Wang}}{1987}]{cho87}
\begin{barticle}
\bauthor{\bsnm{{Chou}}, \binits{D.}},
\bauthor{\bsnm{{Wang}}, \binits{H.}}:
\byear{1987},
\batitle{{The separation velocity of emerging magnetic flux}}.
\bjtitle{\solphys}
\bvolume{110},
\bfpage{81}\,--\,\blpage{99}.
doi:\doiurl{10.1007/BF00148204}.
\end{barticle}
\endbibitem

\bibitem[\protect\citeauthoryear{{Fan}}{2009}]{fan09}
\begin{barticle}
\bauthor{\bsnm{{Fan}}, \binits{Y.}}:
\byear{2009},
\batitle{{Magnetic Fields in the Solar Convection Zone}}.
\bjtitle{Living Reviews in Solar Physics}
\bvolume{6},
\bfpage{4}.
\end{barticle}
\endbibitem

\bibitem[\protect\citeauthoryear{{Frazier}}{1972}]{fra72}
\begin{barticle}
\bauthor{\bsnm{{Frazier}}, \binits{E.N.}}:
\byear{1972},
\batitle{{The Magnetic Structure of Arch Filament Systems}}.
\bjtitle{\solphys}
\bvolume{26},
\bfpage{130}\,--\,\blpage{141}.
doi:\doiurl{10.1007/BF00155113}.
\end{barticle}
\endbibitem

\bibitem[\protect\citeauthoryear{{Gopasyuk}}{1967}]{gop67}
\begin{barticle}
\bauthor{\bsnm{{Gopasyuk}}, \binits{S.I.}}:
\byear{1967},
\batitle{{The velocity field in an active region at spot appearance stag.}}
\bjtitle{Izvestiya Ordena Trudovogo Krasnogo Znameni Krymskoj Astrofizicheskoj
  Observatorii}
\bvolume{37},
\bfpage{29}\,--\,\blpage{43}.
\end{barticle}
\endbibitem

\bibitem[\protect\citeauthoryear{{Gopasyuk}}{1969}]{gop69}
\begin{barticle}
\bauthor{\bsnm{{Gopasyuk}}, \binits{S.I.}}:
\byear{1969},
\batitle{{The velocity field on the two levels in the active region of July
  1966.}}
\bjtitle{Izvestiya Ordena Trudovogo Krasnogo Znameni Krymskoj Astrofizicheskoj
  Observatorii}
\bvolume{40},
\bfpage{111}\,--\,\blpage{126}.
\end{barticle}
\endbibitem

\bibitem[\protect\citeauthoryear{{Grigor'ev}, {Ermakova}, and
  {Khlystova}}{2007}]{gri07}
\begin{barticle}
\bauthor{\bsnm{{Grigor'ev}}, \binits{V.M.}},
\bauthor{\bsnm{{Ermakova}}, \binits{L.V.}},
\bauthor{\bsnm{{Khlystova}}, \binits{A.I.}}:
\byear{2007},
\batitle{{Dynamics of line-of-sight velocities and magnetic field in the solar
  photosphere during the formation of the large active region NOAA 10488}}.
\bjtitle{Astronomy Letters}
\bvolume{33},
\bfpage{766}\,--\,\blpage{770}.
doi:\doiurl{10.1134/S1063773707110072}.
\end{barticle}
\endbibitem

\bibitem[\protect\citeauthoryear{{Grigor'ev}, {Ermakova}, and
  {Khlystova}}{2009}]{gri09}
\begin{barticle}
\bauthor{\bsnm{{Grigor'ev}}, \binits{V.M.}},
\bauthor{\bsnm{{Ermakova}}, \binits{L.V.}},
\bauthor{\bsnm{{Khlystova}}, \binits{A.I.}}:
\byear{2009},
\batitle{{Emergence of magnetic flux at the solar surface and the origin of
  active regions}}.
\bjtitle{Astron. Rep.}
\bvolume{53},
\bfpage{869}\,--\,\blpage{878}.
doi:\doiurl{10.1134/S1063772909090108}.
\end{barticle}
\endbibitem

\bibitem[\protect\citeauthoryear{{Grigor'ev}, {Ermakova}, and
  {Khlystova}}{2011}]{gri11}
\begin{barticle}
\bauthor{\bsnm{{Grigor'ev}}, \binits{V.M.}},
\bauthor{\bsnm{{Ermakova}}, \binits{L.V.}},
\bauthor{\bsnm{{Khlystova}}, \binits{A.I.}}:
\byear{2011},
\batitle{{The dynamics of photospheric line-of-sight velocities in emerging
  active regions}}.
\bjtitle{Astronomy Reports}
\bvolume{55},
\bfpage{163}\,--\,\blpage{173}.
doi:\doiurl{10.1134/S1063772911020041}.
\end{barticle}
\endbibitem

\bibitem[\protect\citeauthoryear{{Guglielmino} \textit{et~al.}}{2006}]{gug06}
\begin{barticle}
\bauthor{\bsnm{{Guglielmino}}, \binits{S.L.}},
\bauthor{\bsnm{{Mart{\'{\i}}nez Pillet}}, \binits{V.}},
\bauthor{\bsnm{{Ruiz Cobo}}, \binits{B.}},
\bauthor{\bsnm{{Zuccarello}}, \binits{F.}},
\bauthor{\bsnm{{Lites}}, \binits{B.W.}}:
\byear{2006},
\batitle{{A Detailed Analysis of an Ephemeral Region .}}
\bjtitle{Memorie della Societa Astronomica Italiana Supplementi}
\bvolume{9},
\bfpage{103}\,--\,\blpage{105}.
\end{barticle}
\endbibitem

\bibitem[\protect\citeauthoryear{{Hagenaar}}{2001}]{hag01}
\begin{barticle}
\bauthor{\bsnm{{Hagenaar}}, \binits{H.J.}}:
\byear{2001},
\batitle{{Ephemeral Regions on a Sequence of Full-Disk Michelson Doppler Imager
  Magnetograms}}.
\bjtitle{\apj}
\bvolume{555},
\bfpage{448}\,--\,\blpage{461}.
doi:\doiurl{10.1086/321448}.
\end{barticle}
\endbibitem

\bibitem[\protect\citeauthoryear{{Harvey} and {Martin}}{1973}]{har73}
\begin{barticle}
\bauthor{\bsnm{{Harvey}}, \binits{K.L.}},
\bauthor{\bsnm{{Martin}}, \binits{S.F.}}:
\byear{1973},
\batitle{{Ephemeral Active Regions}}.
\bjtitle{\solphys}
\bvolume{32},
\bfpage{389}\,--\,\blpage{402}.
doi:\doiurl{10.1007/BF00154951}.
\end{barticle}
\endbibitem

\bibitem[\protect\citeauthoryear{{Kawaguchi} and {Kitai}}{1976}]{kaw76}
\begin{barticle}
\bauthor{\bsnm{{Kawaguchi}}, \binits{I.}},
\bauthor{\bsnm{{Kitai}}, \binits{R.}}:
\byear{1976},
\batitle{{The velocity field associated with the birth of sunspots}}.
\bjtitle{\solphys}
\bvolume{46},
\bfpage{125}\,--\,\blpage{135}.
doi:\doiurl{10.1007/BF00157559}.
\end{barticle}
\endbibitem

\bibitem[\protect\citeauthoryear{{Khlystova}}{2011}]{khl11}
\begin{barticle}
\bauthor{\bsnm{{Khlystova}}, \binits{A.}}:
\byear{2011},
\batitle{{Center-limb dependence of photospheric velocities in regions of
  emerging magnetic fields on the Sun}}.
\bjtitle{\aap}
\bvolume{528},
\bfpage{A7}.
doi:\doiurl{10.1051/0004-6361/201015765}.
\end{barticle}
\endbibitem

\bibitem[\protect\citeauthoryear{{Khlystova}}{2012}]{khl12}
\begin{botherref}
\oauthor{\bsnm{{Khlystova}}, \binits{A.}}:
2012,
{The Relationship between Plasma Flow Velocities and Magnetic Field Parameters
  During the Emergence of Active Regions at the Solar Photospheric Level}.
\textit{\solphys, in Topical Issue ``Advances of European Solar Physics''}.
doi:\doiurl{10.1007/s11207-012-0193-4}.
\end{botherref}
\endbibitem

\bibitem[\protect\citeauthoryear{{Kozu}, {Kitai}, and
  {Funakoshi}}{2005}]{koz05}
\begin{barticle}
\bauthor{\bsnm{{Kozu}}, \binits{H.}},
\bauthor{\bsnm{{Kitai}}, \binits{R.}},
\bauthor{\bsnm{{Funakoshi}}, \binits{Y.}}:
\byear{2005},
\batitle{{Development of Real-Time Frame Selector 2 and the Characteristic
  Convective Structure in the Emerging Flux Region}}.
\bjtitle{\pasj}
\bvolume{57},
\bfpage{221}\,--\,\blpage{234}.
\end{barticle}
\endbibitem

\bibitem[\protect\citeauthoryear{{Kozu} \textit{et~al.}}{2006}]{koz06}
\begin{barticle}
\bauthor{\bsnm{{Kozu}}, \binits{H.}},
\bauthor{\bsnm{{Kitai}}, \binits{R.}},
\bauthor{\bsnm{{Brooks}}, \binits{D.H.}},
\bauthor{\bsnm{{Kurokawa}}, \binits{H.}},
\bauthor{\bsnm{{Yoshimura}}, \binits{K.}},
\bauthor{\bsnm{{Berger}}, \binits{T.E.}}:
\byear{2006},
\batitle{{Horizontal and Vertical Flow Structure in Emerging Flux Regions}}.
\bjtitle{\pasj}
\bvolume{58},
\bfpage{407}\,--\,\blpage{421}.
\end{barticle}
\endbibitem

\bibitem[\protect\citeauthoryear{{Kubo}, {Shimizu}, and {Lites}}{2003}]{kub03}
\begin{barticle}
\bauthor{\bsnm{{Kubo}}, \binits{M.}},
\bauthor{\bsnm{{Shimizu}}, \binits{T.}},
\bauthor{\bsnm{{Lites}}, \binits{B.W.}}:
\byear{2003},
\batitle{{The Evolution of Vector Magnetic Fields in an Emerging Flux Region}}.
\bjtitle{\apj}
\bvolume{595},
\bfpage{465}\,--\,\blpage{482}.
doi:\doiurl{10.1086/377333}.
\end{barticle}
\endbibitem

\bibitem[\protect\citeauthoryear{{Lagg} \textit{et~al.}}{2007}]{lag07}
\begin{barticle}
\bauthor{\bsnm{{Lagg}}, \binits{A.}},
\bauthor{\bsnm{{Woch}}, \binits{J.}},
\bauthor{\bsnm{{Solanki}}, \binits{S.K.}},
\bauthor{\bsnm{{Krupp}}, \binits{N.}}:
\byear{2007},
\batitle{{Supersonic downflows in the vicinity of a growing pore. Evidence of
  unresolved magnetic fine structure at chromospheric heights}}.
\bjtitle{\aap}
\bvolume{462},
\bfpage{1147}\,--\,\blpage{1155}.
doi:\doiurl{10.1051/0004-6361:20054700}.
\end{barticle}
\endbibitem

\bibitem[\protect\citeauthoryear{{Lites}, {Skumanich}, and {Martinez
  Pillet}}{1998}]{lit98}
\begin{barticle}
\bauthor{\bsnm{{Lites}}, \binits{B.W.}},
\bauthor{\bsnm{{Skumanich}}, \binits{A.}},
\bauthor{\bsnm{{Martinez Pillet}}, \binits{V.}}:
\byear{1998},
\batitle{{Vector magnetic fields of emerging solar flux. I. Properties at the
  site of emergence}}.
\bjtitle{\aap}
\bvolume{333},
\bfpage{1053}\,--\,\blpage{1068}.
\end{barticle}
\endbibitem

\bibitem[\protect\citeauthoryear{{Luoni} \textit{et~al.}}{2011}]{luo11}
\begin{barticle}
\bauthor{\bsnm{{Luoni}}, \binits{M.L.}},
\bauthor{\bsnm{{D{\'e}moulin}}, \binits{P.}},
\bauthor{\bsnm{{Mandrini}}, \binits{C.H.}},
\bauthor{\bsnm{{van Driel-Gesztelyi}}, \binits{L.}}:
\byear{2011},
\batitle{{Twisted Flux Tube Emergence Evidenced in Longitudinal Magnetograms:
  Magnetic Tongues}}.
\bjtitle{\solphys}
\bvolume{270},
\bfpage{45}\,--\,\blpage{74}.
doi:\doiurl{10.1007/s11207-011-9731-8}.
\end{barticle}
\endbibitem

\bibitem[\protect\citeauthoryear{{Murray} \textit{et~al.}}{2006}]{mur06}
\begin{barticle}
\bauthor{\bsnm{{Murray}}, \binits{M.J.}},
\bauthor{\bsnm{{Hood}}, \binits{A.W.}},
\bauthor{\bsnm{{Moreno-Insertis}}, \binits{F.}},
\bauthor{\bsnm{{Galsgaard}}, \binits{K.}},
\bauthor{\bsnm{{Archontis}}, \binits{V.}}:
\byear{2006},
\batitle{{3D simulations identifying the effects of varying the twist and field
  strength of an emerging flux tube}}.
\bjtitle{\aap}
\bvolume{460},
\bfpage{909}\,--\,\blpage{923}.
doi:\doiurl{10.1051/0004-6361:20065950}.
\end{barticle}
\endbibitem

\bibitem[\protect\citeauthoryear{{Otsuji} \textit{et~al.}}{2011}]{ots11}
\begin{barticle}
\bauthor{\bsnm{{Otsuji}}, \binits{K.}},
\bauthor{\bsnm{{Kitai}}, \binits{R.}},
\bauthor{\bsnm{{Ichimoto}}, \binits{K.}},
\bauthor{\bsnm{{Shibata}}, \binits{K.}}:
\byear{2011},
\batitle{{Statistical Study on the Nature of Solar-Flux Emergence}}.
\bjtitle{\pasj}
\bvolume{63},
\bfpage{1047}\,--\,\blpage{1057}.
\end{barticle}
\endbibitem

\bibitem[\protect\citeauthoryear{{Pevtsov} and {Lamb}}{2006}]{pev06}
\begin{bchapter}
\bauthor{\bsnm{{Pevtsov}}, \binits{A.}},
\bauthor{\bsnm{{Lamb}}, \binits{J.B.}}:
\byear{2006},
\bctitle{{Plasma Flows in Emerging Sunspots in Pictures}}.
In: \beditor{\bsnm{{Leibacher}}, \binits{J.}},
\beditor{\bsnm{{Stein}}, \binits{R.F.}},
\beditor{\bsnm{{Uitenbroek}}, \binits{H.}} (eds.)
\bbtitle{Solar MHD Theory and Observations: A High Spatial Resolution
  Perspective},
\bsertitle{Astronomical Society of the Pacific Conference Series}
\bseriesno{354},
\bfpage{249}\,--\,\blpage{255}.
\end{bchapter}
\endbibitem

\bibitem[\protect\citeauthoryear{{Scherrer} \textit{et~al.}}{1995}]{sch95}
\begin{barticle}
\bauthor{\bsnm{{Scherrer}}, \binits{P.H.}},
\bauthor{\bsnm{{Bogart}}, \binits{R.S.}},
\bauthor{\bsnm{{Bush}}, \binits{R.I.}},
\bauthor{\bsnm{{Hoeksema}}, \binits{J.T.}},
\bauthor{\bsnm{{Kosovichev}}, \binits{A.G.}},
\bauthor{\bsnm{{Schou}}, \binits{J.}},
\bauthor{\bsnm{{Rosenberg}}, \binits{W.}},
\bauthor{\bsnm{{Springer}}, \binits{L.}},
\bauthor{\bsnm{{Tarbell}}, \binits{T.D.}},
\bauthor{\bsnm{{Title}}, \binits{A.}},
\bauthor{\bsnm{{Wolfson}}, \binits{C.J.}},
\bauthor{\bsnm{{Zayer}}, \binits{I.}},
\bauthor{\bsnm{{MDI Engineering Team}}}:
\byear{1995},
\batitle{{The Solar Oscillations Investigation - Michelson Doppler Imager}}.
\bjtitle{\solphys}
\bvolume{162},
\bfpage{129}\,--\,\blpage{188}.
doi:\doiurl{10.1007/BF00733429}.
\end{barticle}
\endbibitem

\bibitem[\protect\citeauthoryear{{Schoolman}}{1973}]{sch73}
\begin{barticle}
\bauthor{\bsnm{{Schoolman}}, \binits{S.A.}}:
\byear{1973},
\batitle{{Videomagnetograph Studies of Solar Magnetic Fields. II: Field Changes
  in an Active Region}}.
\bjtitle{\solphys}
\bvolume{32},
\bfpage{379}\,--\,\blpage{388}.
doi:\doiurl{10.1007/BF00154950}.
\end{barticle}
\endbibitem

\bibitem[\protect\citeauthoryear{{Shibata} \textit{et~al.}}{1990}]{shi90}
\begin{barticle}
\bauthor{\bsnm{{Shibata}}, \binits{K.}},
\bauthor{\bsnm{{Nozawa}}, \binits{S.}},
\bauthor{\bsnm{{Matsumoto}}, \binits{R.}},
\bauthor{\bsnm{{Sterling}}, \binits{A.C.}},
\bauthor{\bsnm{{Tajima}}, \binits{T.}}:
\byear{1990},
\batitle{{Emergence of solar magnetic flux from the convection zone into the
  photosphere and chromosphere}}.
\bjtitle{\apjl}
\bvolume{351},
\bfpage{L25}\,--\,\blpage{L28}.
doi:\doiurl{10.1086/185671}.
\end{barticle}
\endbibitem

\bibitem[\protect\citeauthoryear{{Sigwarth}, {Schmidt}, and
  {Schuessler}}{1998}]{sig98}
\begin{barticle}
\bauthor{\bsnm{{Sigwarth}}, \binits{M.}},
\bauthor{\bsnm{{Schmidt}}, \binits{W.}},
\bauthor{\bsnm{{Schuessler}}, \binits{M.}}:
\byear{1998},
\batitle{{Upwelling in a young sunspot}}.
\bjtitle{\aap}
\bvolume{339},
\bfpage{L53}\,--\,\blpage{L56}.
\end{barticle}
\endbibitem

\bibitem[\protect\citeauthoryear{{Snodgrass}}{1983}]{sno83}
\begin{barticle}
\bauthor{\bsnm{{Snodgrass}}, \binits{H.B.}}:
\byear{1983},
\batitle{{Magnetic rotation of the solar photosphere}}.
\bjtitle{\apj}
\bvolume{270},
\bfpage{288}\,--\,\blpage{299}.
doi:\doiurl{10.1086/161121}.
\end{barticle}
\endbibitem

\bibitem[\protect\citeauthoryear{{Solanki} \textit{et~al.}}{2003}]{sol03}
\begin{barticle}
\bauthor{\bsnm{{Solanki}}, \binits{S.K.}},
\bauthor{\bsnm{{Lagg}}, \binits{A.}},
\bauthor{\bsnm{{Woch}}, \binits{J.}},
\bauthor{\bsnm{{Krupp}}, \binits{N.}},
\bauthor{\bsnm{{Collados}}, \binits{M.}}:
\byear{2003},
\batitle{{Three-dimensional magnetic field topology in a region of solar
  coronal heating}}.
\bjtitle{\nat}
\bvolume{425},
\bfpage{692}\,--\,\blpage{695}.
doi:\doiurl{10.1038/nature02035}.
\end{barticle}
\endbibitem

\bibitem[\protect\citeauthoryear{{Strous} and {Zwaan}}{1999}]{str99}
\begin{barticle}
\bauthor{\bsnm{{Strous}}, \binits{L.H.}},
\bauthor{\bsnm{{Zwaan}}, \binits{C.}}:
\byear{1999},
\batitle{{Phenomena in an Emerging Active Region. II. Properties of the Dynamic
  Small-Scale Structure}}.
\bjtitle{\apj}
\bvolume{527},
\bfpage{435}\,--\,\blpage{444}.
doi:\doiurl{10.1086/308071}.
\end{barticle}
\endbibitem

\bibitem[\protect\citeauthoryear{{Strous} \textit{et~al.}}{1996}]{str96}
\begin{barticle}
\bauthor{\bsnm{{Strous}}, \binits{L.H.}},
\bauthor{\bsnm{{Scharmer}}, \binits{G.}},
\bauthor{\bsnm{{Tarbell}}, \binits{T.D.}},
\bauthor{\bsnm{{Title}}, \binits{A.M.}},
\bauthor{\bsnm{{Zwaan}}, \binits{C.}}:
\byear{1996},
\batitle{{Phenomena in an emerging active region. I. Horizontal dynamics.}}
\bjtitle{\aap}
\bvolume{306},
\bfpage{947}\,--\,\blpage{959}.
\end{barticle}
\endbibitem

\bibitem[\protect\citeauthoryear{{Tarbell} \textit{et~al.}}{1989}]{tar89}
\begin{bchapter}
\bauthor{\bsnm{{Tarbell}}, \binits{T.D.}},
\bauthor{\bsnm{{Topka}}, \binits{K.}},
\bauthor{\bsnm{{Ferguson}}, \binits{S.}},
\bauthor{\bsnm{{Frank}}, \binits{Z.}},
\bauthor{\bsnm{{Title}}, \binits{A.M.}}:
\byear{1989},
\bctitle{{High - resolution observations of emerging magnetic flux}}.
In: \beditor{\bsnm{{O.~von der Luehe}}} (ed.)
\bbtitle{High spatial resolution solar observations},
\bfpage{506}\,--\,\blpage{520}.
\end{bchapter}
\endbibitem

\bibitem[\protect\citeauthoryear{{Toriumi} and {Yokoyama}}{2010}]{tor10}
\begin{barticle}
\bauthor{\bsnm{{Toriumi}}, \binits{S.}},
\bauthor{\bsnm{{Yokoyama}}, \binits{T.}}:
\byear{2010},
\batitle{{Two-step Emergence of the Magnetic Flux Sheet from the Solar
  Convection Zone}}.
\bjtitle{\apj}
\bvolume{714},
\bfpage{505}\,--\,\blpage{516}.
doi:\doiurl{10.1088/0004-637X/714/1/505}.
\end{barticle}
\endbibitem

\bibitem[\protect\citeauthoryear{{Toriumi} \textit{et~al.}}{2011}]{tor11}
\begin{barticle}
\bauthor{\bsnm{{Toriumi}}, \binits{S.}},
\bauthor{\bsnm{{Miyagoshi}}, \binits{T.}},
\bauthor{\bsnm{{Yokohama}}, \binits{T.}},
\bauthor{\bsnm{{Isobe}}, \binits{H.}},
\bauthor{\bsnm{{Shibata}}, \binits{K.}}:
\byear{2011},
\batitle{{Dependence of the Magnetic Energy of Solar Active Regions on the
  Twist Intensity of the Initial Flux Tubes}}.
\bjtitle{\pasj}
\bvolume{63},
\bfpage{407}\,--\,\blpage{415}.
\end{barticle}
\endbibitem

\bibitem[\protect\citeauthoryear{{van Driel-Gesztelyi} and
  {Petrovay}}{1990}]{dri90}
\begin{barticle}
\bauthor{\bsnm{{van Driel-Gesztelyi}}, \binits{L.}},
\bauthor{\bsnm{{Petrovay}}, \binits{K.}}:
\byear{1990},
\batitle{{Asymmetric flux loops in active regions.}}
\bjtitle{\solphys}
\bvolume{126},
\bfpage{285}\,--\,\blpage{298}.
doi:\doiurl{10.1007/BF00153051}.
\end{barticle}
\endbibitem

\bibitem[\protect\citeauthoryear{{Xu}, {Lagg}, and {Solanki}}{2010}]{xu10}
\begin{barticle}
\bauthor{\bsnm{{Xu}}, \binits{Z.}},
\bauthor{\bsnm{{Lagg}}, \binits{A.}},
\bauthor{\bsnm{{Solanki}}, \binits{S.K.}}:
\byear{2010},
\batitle{{Magnetic structures of an emerging flux region in the solar
  photosphere and chromosphere}}.
\bjtitle{\aap}
\bvolume{520},
\bfpage{A77}.
doi:\doiurl{10.1051/0004-6361/200913227}.
\end{barticle}
\endbibitem

\bibitem[\protect\citeauthoryear{{Zwaan}, {Brants}, and {Cram}}{1985}]{zwa85}
\begin{barticle}
\bauthor{\bsnm{{Zwaan}}, \binits{C.}},
\bauthor{\bsnm{{Brants}}, \binits{J.J.}},
\bauthor{\bsnm{{Cram}}, \binits{L.E.}}:
\byear{1985},
\batitle{{High-resolution spectroscopy of active regions. I - Observing
  procedures}}.
\bjtitle{\solphys}
\bvolume{95},
\bfpage{3}\,--\,\blpage{14}.
doi:\doiurl{10.1007/BF00162632}.
\end{barticle}
\endbibitem

\end{thebibliography}

\end{article}
\end{document}